\documentclass[aps,preprintnumbers,superscriptaddress,twocolumn,amsmath,amssymb,floatfix,prl]{revtex4} 
\pdfoutput=1
\usepackage{graphicx}
\usepackage[usenames,dvipsnames]{color}
\usepackage[colorlinks,bookmarks=false,citecolor=NavyBlue,linkcolor=Red,urlcolor=blue]{hyperref}
\usepackage[export]{adjustbox}

\newcommand\be{\begin{equation}}
\newcommand\bea{\begin{eqnarray}}
\newcommand\bes{\begin{subequations}}
\newcommand\esu{\end{subequations}}
\newcommand\ee{\end{equation}}
\newcommand\eea{\end{eqnarray}}

\newcommand{\cmmnt}[1]{}

\def\doi{http://dx.doi.org/}

\usepackage{braket}

\newcommand\ocite[1]{[\onlinecite{#1}]}

\begin{document}

\title{Non-Equilibrium Steady State generated by a moving defect:\\
the supersonic threshold }

\author{Alvise Bastianello}
\affiliation{SISSA \& INFN, via Bonomea 265, 34136 Trieste, Italy}

\author{Andrea De Luca}
\affiliation{The Rudolf Peierls Centre for Theoretical Physics, Oxford University, Oxford, OX1 3NP, United Kingdom}

\begin{abstract}
We consider the dynamics of a system of free fermions on a 1D lattice in the presence of a defect 
moving at constant velocity. 
The defect has the form of a localized time-dependent variation of the chemical potential and induces 
at long times a Non-Equilibrium Steady State (NESS), which spreads around the defect. 
We present a general formulation which allows recasting the time-dependent 
protocol in a scattering problem on a static potential.
We obtain a complete characterization of the NESS.
In particular, we show a strong dependence on the defect velocity and the existence 
of a sharp threshold when such velocity exceeds the speed of sound. 
Beyond this value, the NESS is not produced and remarkably 
the defect travels without significantly perturbing the system.
We present an exact solution for a $\delta-$like defect traveling with an arbitrary velocity and
we develop a semiclassical approximation which provides accurate results for smooth defects.
\end{abstract}


\maketitle

Recent experimental advances  in the context of cold atoms \cite{exp1,exp2,exp3,exp4,exp5,exp6,exp7,exp8,exp9,exp10,exp11,newExp1,newExp2,newExp3,newExp4,newExp5} converted the study of out-of-equilibrium closed quantum systems from an academic 
debate to a concrete and extremely active research topic.
In this context, the simplest protocol 
is known as \emph{quantum quench}~\cite{calabrese-cardy},
where the system is brought out-of-equilibrium by a sudden change of a coupling constant.
Despite the unitary time evolution, in the thermodynamic limit,
local observables reach a time-independent expectation value and 
the system locally equilibrates \cite{pssv}.
{One dimensional systems} have played a special role 
because of the presence of special techniques, such  
as \emph{conformal field theory} \cite{difrancesco} 
and \emph{integrability} \cite{Korepin,smirnov}.
In particular, in \emph{homogeneous quenches} 
a global parameter is modified and several studies with a number of 
integrable models and initial conditions~\cite{ggew1,ggew2,ggew3,ggew4,ggew5,ggew6,ggew7,ggew8,ggew9,ggew10} 
(see also \cite{specialissue} for a review) have confirmed 
the validity of the Generalized Gibbs Ensemble (GGE)~\cite{gge1}.
The GGE hypothesis prescribes that the steady state assumes a thermal form, 
but with an extended set of temperatures, conjugated to each (quasi) local 
conserved quantity present in the model~\cite{gge1,gge3}.
The value of such temperatures is then fixed by 
the initial expectation value of conserved quantities~\cite{ggef1,ggef2,ggef3,ggef4,lch1,lch2,lch3,lch4,lch5,lch6,lchf1,lchf2,lchf3,lchf4,lchf5}.

A more complex scenario emerges for \emph{inhomogeneous quenches}, 
where, because of an asymmetry in the initial state or in the final Hamiltonian,
translational invariance is explicitly broken. The simplest case is the one in which a localized 
defect perturbs an otherwise homogeneous system.
As the spreading of correlations is bounded by a maximal speed of sound $v_s$,
this defect cannot affect immediately the whole (thermodynamically large) system~\cite{liebrob, localquenches}. 
In the presence of \emph{ballistic} dynamics, at late times $t$ and large distance $x$ from the defect, 
the system reaches a Locally Quasi-Stationary State (LQSS)~\cite{transportbertini}, whose properties depend 
only on the ray $\zeta=x/t$ inside the lightcone $|\zeta| < v_s$.
The infinite time limit at finite distance (i.e. $\zeta=0$) corresponds instead to the Non-Equilibrium Steady State (NESS).
Our present understanding of LQSS is based on numerical studies \cite{NESSnum1,NESSnum2,NESSnum3}, free models~\cite{NESSf0,NESSf1,NESSf2,NESSf3,NESSf4,
NESSf5,NESSf6,NESSf7,NESSf8,NESSf9,NESSf10,NESSf11,NESSf12}, CFT \cite{NESSc1,NESSc2,NESSc3,NESSc4,NESSc5,PB} 
and only recently on truly interacting integrable models \cite{NESSI1,NESSI2,NESSI3,transportbertini,hydrodoyon1,PBCF}. 
In particular, the hydrodynamic description \cite{transportbertini, hydrodoyon1,hydrodoyon2,PBCF} 
has led to exact results with possible applications to several contexts \cite{IlNard,new_doyon1,new_doyon2,xxzhydro,karrasch,doyonspohn}.

In this Letter, we consider instead a \emph{moving defect}. This setting offers an additional parameter to control the stationary state, particularly interesting in those systems whose excitation possess a maximum velocity. 
 Moving impurities have already been experimentally probed, in particular in \ocite{newExp4}, the case of a Tonks Girardeau gas\, \cite{hardcore} (intimately linked to the free fermion case here analyzed) was considered.
While several works have considered moving impurities in several contexts \ocite{refB1,refB2,refB3,refB4,imp1,imp2,imp3,imp4,imp5,imp6,imp7}, the long-time out-of-equilibrium dynamics has never been addressed so far. In particular, does the system still reacts forming a LQSS? How the LQSS changes for different velocities of the defect? 
We explore these questions in the prototypical case of hopping fermions on a lattice, which is amenable to a full analytical treatment
still retaining a rich phenomenology. 
At time $t>0$, the dynamics is governed by the following time dependent Hamiltonian:
\be
H=\sum_{j}-\frac{1}{2}\left(d^\dagger_{j}d_{j+1}+d^\dagger_{j+1}d_j\right)+V(j-vt)d^\dagger_jd_j\, .\label{latHam}
\ee
where $d_j$ are the fermionic operators $\{d_j,d_l^\dagger\}=\delta_{j,l}$. With the current choice of couplings, the unperturbed system has a sound velocity $v_s=1$.
This lattice model can be mapped in the XX spin chain \cite{solutionXY1,solutionXY2,solutionXY3}, where the defect plays the role of a traveling magnetic impurity.
The system is assumed to be initially in the unperturbed ground state ($V=0$) at fixed particle number (the finite temperature case being a trivial generalization). 
Here, the impurity is suddenly created and put in motion; other similar settings (e.g. the motion of a pre-existing defect) would lead to the same late time physics.
We show the emergence of a \emph{moving} LQSS, whose rays refer to the instantaneous position of
the defect. The amplitude of the LQSS is showed to be suppressed when the velocity of the defect is increased and the formation of a LQSS becomes \emph{impossible} for a \emph{supersonic} impurity.
We fully determine the \emph{exact} LQSS generated by a $\delta-$like perturbation and provide a semiclassical expression of the LQSS, 
valid for smooth defects.
Even though the model we are considering is free, the emergent LQSS displays absolutely non-trivial features understood thanks to its exact description, while approximated methods commonly used could lead to misleading conclusions (see \ocite{suppl} for the discussion of the Luttinger-Liquid approximation \ocite{LL1,LL2}). 
\
\emph{The LQSS as a scattering problem. ---} 
The effect of the moving impurity is best understood in a pictorial representation where the initial state can be regarded as a gas of excitations. The excitations move freely in the space until the defect is met, then a scattering event takes place and the excitation continues in a free motion, with a different momentum. The non-trivial LQSS is due to the scattered particles, spreading ballistically from the defect. Thereafter, we make this argument rigorous.
The initial state is a filled \emph{Fermi sea}, with gaussian correlations.
As the post-quench Hamiltonian is quadratic, all the local observables at any time are fixed by the two-point correlators via the Wick theorem. 
Thereafter, we focus only on the case $v> 0$. Changing the reference frame to set the defect at rest would remove the explicit time-dependence of the problem, but such a program is foiled by the discreteness of the lattice.
This difficulty can be circumvented through a map to a continuous fermionic model $\{c_{x},c^\dagger_{y}\}=\delta(x-y)$ with Hamiltonian
\be
H_\text{c}=\int dx\, -\frac{1}{2}\left(c^\dagger_{x}c_{x+1}+c^\dagger_{x+1}c_x\right)+V(x-vt)c^\dagger_xc_x\, .\label{contHam}
\ee
From this model all the discrete correlation functions are  \emph{exactly} recovered. Indeed,
$H_{\text{c}}$ only couples a coordinate $x$ with $x+n$, $n\in \mathbb{Z}$.
On this sublattice, continuous and discrete (normal ordered) correlation functions satisfy the same equation of motion, thus their solution is the same provided consistent initial conditions $\langle c^\dagger_j c_l\rangle_{t=0}=\langle d^\dagger_j d_l\rangle_{t=0}$ have been chosen.
We can thus employ \eqref{contHam} to study the dynamics of the system and later restrict ourselves to integer positions. 
 This approach leaves us the freedom of arbitrarily choosing the correlator at non-integer values; a convenient choice is to assume the initial state in the momentum space is described by the same Fermi sea of the discrete model.
As $x$ is a continous coordinate, we can now introduce a reference frame $c_x=\tilde{c}_{x-vt}$ 
where the defect is \emph{at rest}. In terms of this new field, the equation of motion can be derived from the time-independent Hamiltonian
\be
\tilde{H}_\text{c}=\int dx\,iv\tilde{c}^\dagger_x\partial_x \tilde{c}_x -\frac{1}{2}\left(\tilde{c}^\dagger_{x}\tilde{c}_{x+1}+\tilde{c}^\dagger_{x+1}\tilde{c}_x\right)+V(x)\tilde{c}^\dagger_x\tilde{c}_x\, .\label{contdefHam} 
\ee
Clearly, the dynamics induced by \eqref{contdefHam} depends on the details of $V(x)$.
\begin{figure}[t]
\begin{center}
\includegraphics[width=0.95\columnwidth,valign=l]{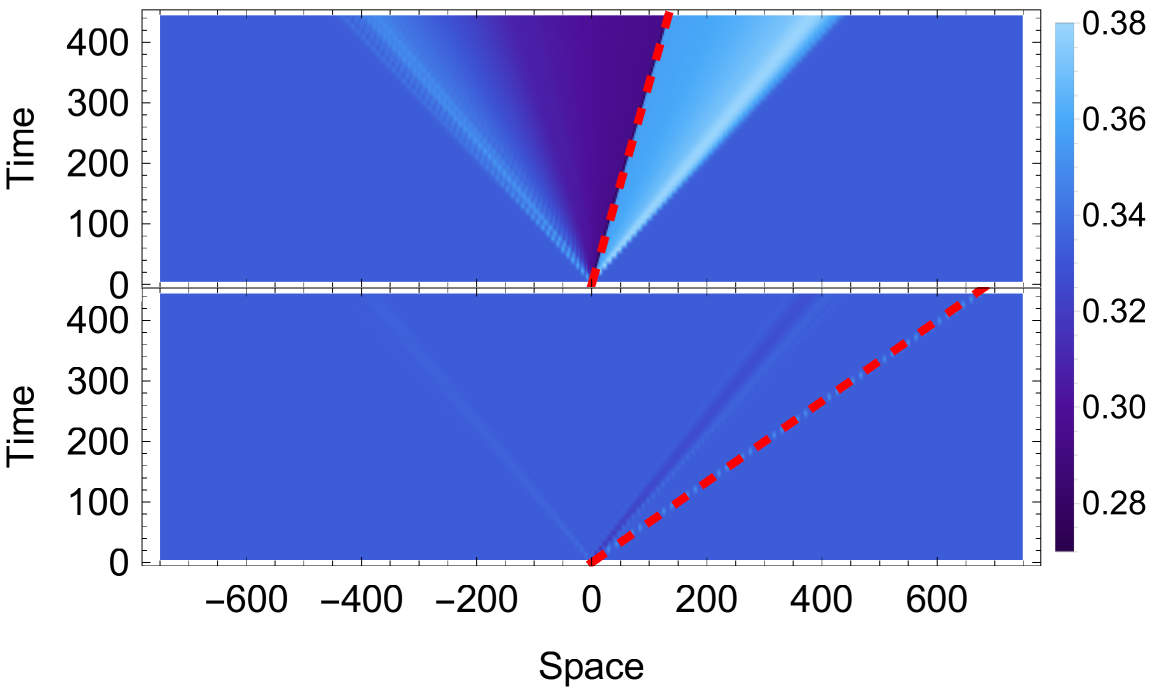}
\caption{\label{fig_cone}Fermion density generated by a $\delta-$like defect ($V(x)=c\delta(x)$, $c=0.5$) moving at $v=0.3$ 
(top) and $v=1.5$ (bottom). The defect is initially placed at zero and it moves along the line dashed in red. 
In the subsonic case the constant values along the rays indicate the realization of a LQSS, 
which is instead absent in the supersonic regime.}
\end{center}
\end{figure}
However, being the defect localized, we can use scattering theory. 
We introduce the mode operators $\eta_k=\int dx\, \psi^*_k(x)\tilde{c}_x$ 
where the $\psi_k(x)$ is the normalized wavefunction satisfying the Lippmann-Schwinger equation \cite{sakurai}.
In other words, far away from the defect, it assumes the simple form of a scattering problem
\footnote{We ignore the possibility of bound states since they are completely irrelevant in the study 
of the LQSS, concerning the latter the scaling region far from the defect.}
\be
\psi_k(x)=\theta(-xv(k))\frac{e^{ikx}}{\sqrt{2\pi}}+\sum_{k_n} S_{k\to k_n}\theta(x v(k_n)) \frac{e^{ik_n x}}{\sqrt{2\pi}}\, ,\label{asymp}
\ee
where the incoming wave is expanded into outgoing waves weighted with the scattering amplitudes $S_{k\to k_n}$.
The wavevectors $k_n$ are obtained via energy conservation $E(k)=E(k_n)$, with 
$E(k)=-\cos(k)-vk$ the single-particle energy in the defect reference frame and $v(k)=dE(k)/dk$ its group velocity.
Corrections to (\ref{asymp}) vanish exponentially in the distance and the 
scattering amplitude takes the form
\be
S_{k\to k_n}=\delta_{k,k_n}-2i\pi|v(k_n)|^{-1}\langle k_n|\hat V|\psi_k\rangle\ \label{Smatrix}
\ee

where we introduced a bra/ket notation with $\braket{x|k} = e^{i k x}/\sqrt{2\pi}$ and $\hat V(x')\ket{x} = V(x')\delta(x-x')\ket{x}$. 
The unitarity of the Lippmann-Schwinger equation permits to derive an \emph{exact sum rule} whose explicit derivation is left to \ocite{suppl}.
The initial two point correlator is diagonal in the momentum space, thus we are ultimately led to consider the time-evolution of plane waves $f_k(x,t)=\braket{x|e^{-iHt}|k}$. In terms of the eigenbasis $|\psi_q\rangle$ is
\be
f_k(x,t)=\sqrt{2\pi}\int dq\, e^{-iE(q) t}\psi_q(x)\langle \psi_q| k\rangle\, .
\ee
The large time behavior of $f_k$ is readily extracted using (\ref{asymp}) together with the aforementioned sum rule for $S$, as the corrections to the asymptotic approximation (\ref{asymp}) are ineffective in the LQSS scaling limit \cite{suppl}. 
We can then obtain the LQSS two-point correlator in the form of a ray-dependent GGE with an excitation density $\rho_\zeta(k)$, being the ensemble gaussian and (locally) diagonal in the momentum 
space~\cite{gaussification1,gaussification2,gaussification3,gaussification4}
\begin{eqnarray}
&&\langle \tilde{c}^\dagger_x\tilde{c}_y\rangle_t=\int_{-\pi}^{\pi}\frac{dk}{2\pi}\rho_\zeta(k)e^{ik(y-x)}=\label{corr_LQSS}\\
&&\nonumber\int_{-k_f}^{k_f}\frac{dk}{2\pi} e^{ik(y-x)}\left[1-\theta(\zeta v(k))\theta(|v(k)|-|\zeta|)\right]+\\
\nonumber&&\sum_{k_n}|S_{k\to k_n}|^2\theta(\zeta v(k_n))\theta(|v(k_n)|-|\zeta|)e^{ik_n(y-x)}\, .
\end{eqnarray}
As the two point correlator is a decaying function in the relative distance, 
in the large time limit, one must set $x/t\sim y/t\sim \zeta$. 
Eq.~\eqref{corr_LQSS} is easily interpreted: when a particle of momentum $k$ collides with the defect, it can scatter in 
several channels $k_n$ with amplitudes $S_{k\to k_n}$. This produces a \emph{hole} 
in the old momentum and a flux of particles with the new momenta:
these spread ballistically at their own velocity $v(k_n)$ leading to the ray dependence on $\zeta$.
In Fig. \ref{fig_cone}, we plot the space-time profile of the fermion density for a $\delta-$like defect, 
whose detailed analysis will be considered later.
With no further information on $S$, we can discuss the behavior of the LQSS in terms of $v$ analyzing the scattering channels, identified by $E(k_n)=E(k)$. For generic $v$, several channels are open (a divergent number letting $v\to 0$) and they diminish increasing $v$ until, for a supersonic defect, the excitations are unavoidably purely transmitted. In this case, the sum rule obeyed by $S$ forces $|S_{k\to k}|=1$  \ocite{suppl} (simply interpreted as particle conservation).
Therefore, in the supersonic regime, 
the LQSS can \emph{never} be produced (see also Fig. \ref{fig_cone}). 
Note that the LQSS (\ref{corr_LQSS}) emerges at late times and large distances 
from the defect, when Eq. (\ref{asymp}) holds: despite the absence of LQSS, a supersonic defect still gives non-trivial effects localized on the impurity.
We will further analyze this behavior within the semiclassical approximation.
\begin{figure}[t!]
\advance\leftskip-1cm
\includegraphics[width=0.90\columnwidth]{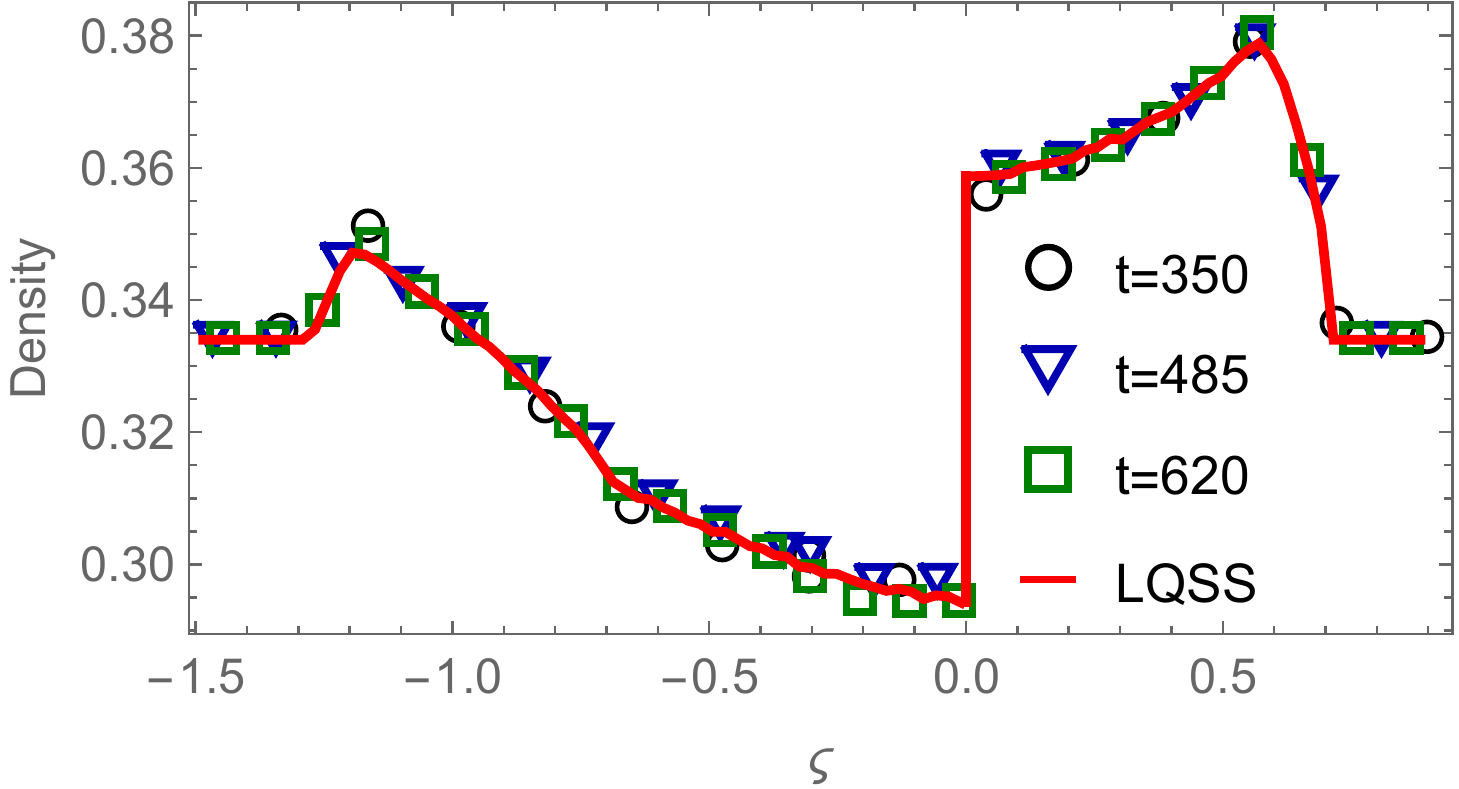}      
\caption{\label{fig1}The numeric fermionic density $\langle d_j^\dag d_j \rangle$ with $j = (v + \zeta) t$ as a function of $\zeta$ at  three different times is tested against the analytic LQSS for 
a $\delta$-like defect $V(x) = c \delta(x)$ with $c=0.5$ and velocity $v=0.3$. The density is fixed by $k_f=\pi/3$. 
The oscillations mentioned in the main text are smeared out by averaging on neighboring sites.
}
\end{figure}
\

\begin{figure}[t!] 
\includegraphics[width=0.98\columnwidth]{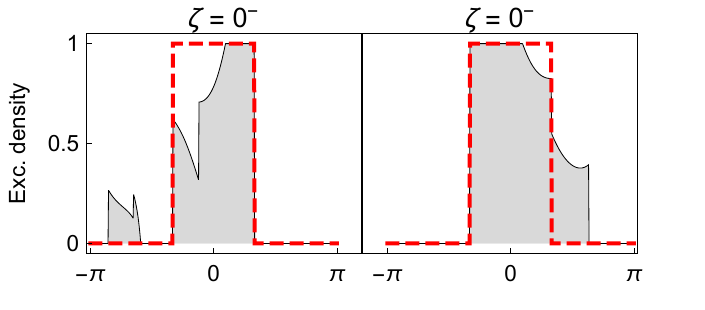}      
\caption{\label{fig_exc}Analytic prediction for the excitation density profile propagating on the left (left panel) and on the right (right panel) of a $\delta-$like defect.
 The dashed red line is the initial Fermi sea, black line the spreading excitation whose integral (shaded area) equals the spatial fermionic density, which jumps around the defect  (see Fig.~\ref{fig1}). The same parameters of Fig.~\ref{fig1} are used.
}
\end{figure}

\emph{An exactly solvable case. ---} We now consider an example of defect
for which the LQSS can be \emph{exactly} determined, i.e. the limit of an extremely narrow defect $V(x)=c\delta(x)$. In the discrete model (\ref{latHam}) the Dirac-$\delta$ is ill-defined when $v=0$.
However, this is not the case for a defect in motion $v\ne 0$: the $\delta-$defect 
represents an impulsive kick traveling along the lattice and leads to well-defined equations of motion. 
The detailed calculations can be found in \cite{suppl}, here we simply report the result. Referring to eq. (\ref{Smatrix}), we have
\be
2\pi \langle k_n|V|\psi_k\rangle=-\Big[\sum_{k_m}\frac{1}{2i|v(k_{m})|}-\frac{1}{2v}\cot\left(\frac{c}{2v}\right)+\mathcal{I}(k)\Big]^{-1}\, .\label{vdelta}
\ee
Above, $\mathcal{I}(k)=\mathcal{P}\int \frac{dq}{2\pi}(E(k)-E(q))^{-1}$, where the principal value prescription in integrating the singular points is assumed.
In Fig. \ref{fig1} the exact solution for the fermion density is tested against numerical simulations. 
Numerical data show persistent oscillations due to the interference of the various scattering channels.  
These oscillations decay far away from the defect and are therefore inessential in the LQSS scaling limit, but are nevertheless captured by scattering theory (see \ocite{suppl}).
The non-trivial density profile is only one of the manifestations of the LQSS, being the complicated structure of the underlying scattering best appreciated in the excitation density propagating from the defect (Fig.~\ref{fig_exc}).
\

\emph{The semiclassical approximation. ---} For general potential, determining the scattering amplitudes requires some approximation schemes. 
Here, we develop a semiclassical analysis in which the scattering interpretation is most easily displayed. 
Our derivation is based on the Wigner distribution \cite{wigner1,wigner2,wigner3}. Semiclassical approaches are commonly found in 
literature \cite{wigners1,wigners2,wigners3,wigners4,wigners5} (see also \ocite{refA1,refA2,refA3,F17} for quantum corrections) even though, at the best of our knowledge, 
the problem at hand has never been addressed.
Consider the two-point correlator $\mathcal{C}_t(x,s)=\langle \tilde{c}^\dagger_{x+s/2}\tilde{c}_{x-s/2}\rangle_t$, 
under the assumption of:
i) weak dependence with respect to the $x$ coordinate; ii) 
fast decay of $\mathcal{C}_t(x,s)$ as a function of $s$, on a length scale much smaller than the length 
on which the defect varies, we can approximate the equation of motions of the correlator as \footnote{The derivative expansion is not completely rigorous for slow decaying correlators (as for the ground state), but a more rigorous and lengthy derivation of the scattering amplitudes from the WKB approximation gives the same result.}
\begin{eqnarray}
\nonumber&&i\partial_t\mathcal{C}_t(x,s)=\frac{1}{2}\partial_x\left(\mathcal{C}_t\left(x,s+1\right)-\mathcal{C}_t\left(x,s-1\right)\right)\\
&&-s\partial_x V(x)\mathcal{C}_t(x,s)+iv\partial_x\mathcal{C}_t(x,s)+...
\end{eqnarray}
\begin{figure}[t]
\includegraphics[width=0.9\columnwidth]{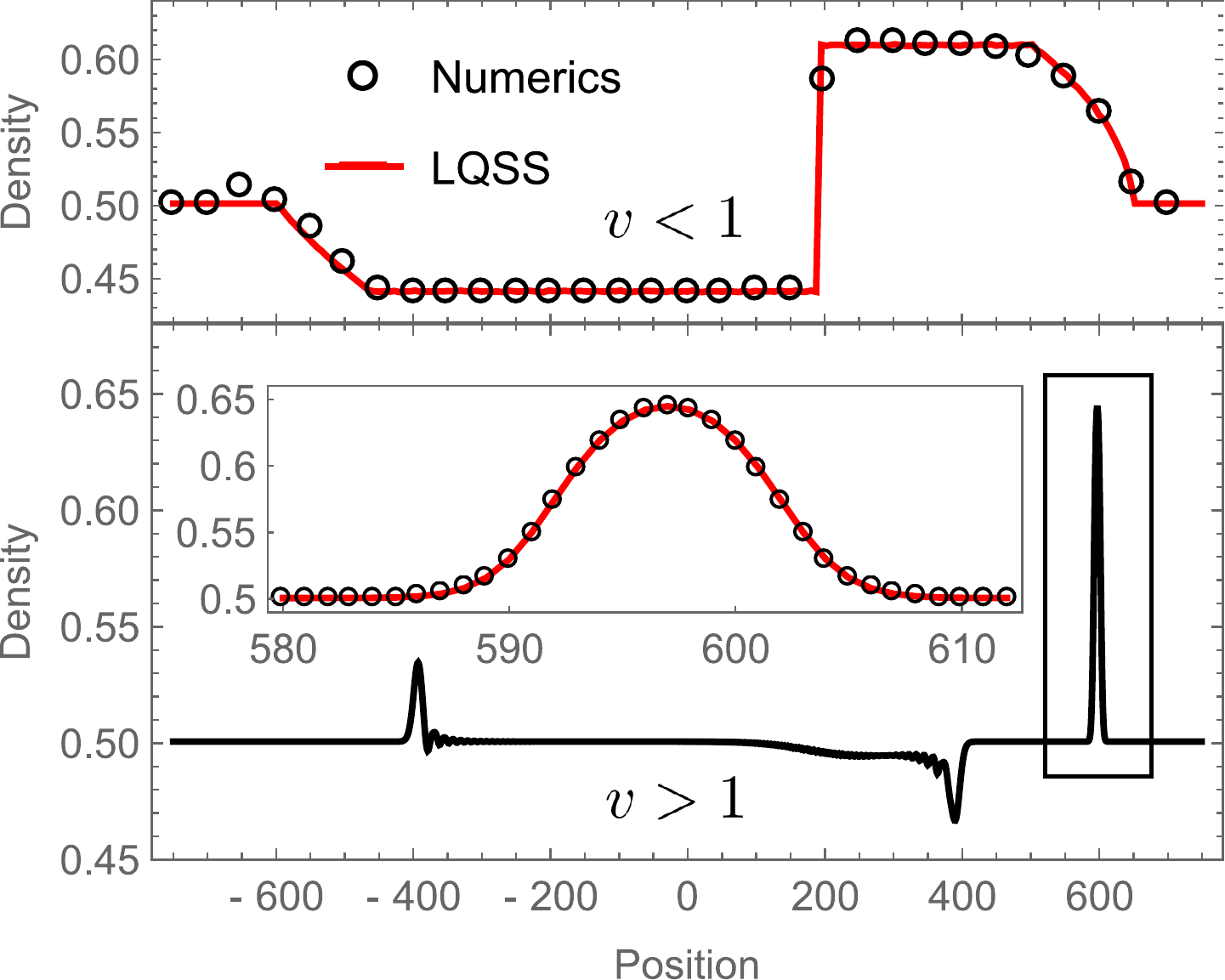}
\caption{\label{fig2}
The numerical fermionic density is tested against the semiclassical LQSS for a gaussian shaped repulsive potential $V(x)=e^{-\sigma x^2}$ with $\sigma=0.04$, $k_f=\pi/2$. Top: subsonic defect ($v=0.3$) and $t=636$.  The defect is placed where the discontinuity occurs and a LQSS is produced. 
Bottom: supersonic defect ($v=1.5$, $t=388$) and placed in correspondence with the rightmost peak: the LQSS is indeed absent and the semiclassical prediction captures the density profile on the defect (inset). The corrections to the analytic prediction are a combined effect of \emph{i)} the particles initially sat 
on the defect that have not yet managed to spread \emph{ii)}  the delay time experienced by the particles scattering on the defect. Both these effects are negligible 
at late time and in the scaling limit. Quantum effects can be recognized in the propagating ripples \ocite{refA1,refA2,refA3}, more evident in the supersonic case. }
\end{figure}
Higher derivatives of the correlators and of the defect are neglected.
Being the initial state homogeneous, this approximation is verified for a smooth potential $V(x)$.
In terms of the Wigner function $\rho_t(x,k)$ defined as
\be
\rho_t(x,k)=\int^\infty_{-\infty} ds \,e^{iks}\,\mathcal{C}_t(x,s)\label{wigner}\;,
\ee
these apparently complicated equations reduce to a \emph{classical Liouville equation} 
\be
\partial_t\rho_t+\{\mathcal{H}_\text{cl},\rho_t\}_{k,x}=0\, ,\hspace{1pc}\mathcal{H}_\text{cl}=-\cos(k)-vk+V(x)\, .
\ee
Above $\{\ldots\}_{k,x}$ indicates the standard Poisson bracket and $\mathcal{H}_\text{cl}$ is the classical Hamiltonian. 
In the semiclassical limit, $\rho_t$ evolves as the density distribution of particles subjected to the classical equations of motion $\dot{x}=v(k)$, $\dot{k}=-\partial_xV(x) $.
The precise solution $x(t), k(t)$ of these equations depends on the details of $V(x)$, but the trajectories $k(x)$ of the particles can 
be easily computed combining the conservation of the classical energy and the determination of the turning points, i.e. 
those points where $dx/dt=0$ (see \ocite{suppl}).
In the semiclassical language, computing the LQSS is reduced to simple classical scattering processes: 
whether a particle is reflected or transmitted by the defect is simply determined reconstructing the trajectories. 
The case of a supersonic defect is remarkably simple, since the equation of motion does not have turning points. In this case we readily obtain
\be
\lim_{t\to\infty}\mathcal{C}_t(x,s)=\int_{-k_f}^{k_f} \frac{dk}{2\pi}\left|\frac{v(k)}{v(q_{k,x})}\right|e^{-iq_{k,x} s}\, ,
\ee
valid also in proximity of the defect (within the semiclassical approximation). Above, $q_{k,x}$ is determined by energy conservation 
\be
-\cos(q_{k,x})-v q_{k,x}+V(x)=-\cos(k)-vk\, .
\ee
which has a unique solution for a supersonic $v$.
The semiclassical results and the numerics are compared in Fig. \ref{fig2}, further details about the semiclassical approximation are left to \cite{suppl}.

\emph{Conclusions and outlook.\,\,\,}We analyzed a local quench where the defect is moving at constant velocity in a system of 
hopping fermions on a lattice. In particular, we focused on the emergence of a Locally-Quasi-Stationary State and we studied 
the dependence of the latter in terms of the velocity of the defect. With general arguments, we showed the impossibility of a 
LQSS formation for supersonic defects. 
We provided \emph{exact} results for a $\delta-$like defect and a 
semiclassical analysis for general shapes.
We are confident that our framework can be studied in forthcoming cold-atoms experiments,
as in Ref. \ocite{newExp4}, a very similar setting was realized. The moving defect would then be a intriguing 
way to induce non-equilibrium dynamics and to probe the scattering properties of quasi-particle excitations.

Being the hopping fermions a free model, its simplicity allowed for exact computations: an  intriguing question concerns the same problem in a truly interacting model like integrable spin chains. 
We expect that a promising investigation line could be the recently introduced generalized hydrodynamics~\cite{transportbertini,hydrodoyon1,hydrodoyon2,IlNard,new_doyon1,new_doyon2,xxzhydro,karrasch,doyonspohn,PBCF},
which for a free model, reduces precisely to the semiclassical approach we used. This will be the subject of a forthcoming publication.
Another natural approach would be the use of the recently introduced curved-CFT formalism \cite{curvecft}.
\ 

\emph{Acknowledgment.\,\,\,}
We are grateful to Mario Collura, Pasquale Calabrese, Fabian Essler and Giuseppe Mussardo for useful discussions.
This work was supported by the EPSRC Quantum Matter in and out of Equilibrium Ref. EP/N01930X/1 (A.D.L.).


\onecolumngrid
\newpage 

\setcounter{equation}{0}            
\setcounter{section}{0}             
\renewcommand\thesection{\Alph{section}}    
\renewcommand\thesubsection{\arabic{subsection}}    
\renewcommand{\thetable}{S\arabic{table}}
\renewcommand{\theequation}{S\arabic{equation}}
\renewcommand{\thefigure}{S\arabic{figure}}

\begin{center}
{\Large Supplementary Material\\ 
Non Equilibrium Steady State generated by a moving defect:\\
the supersonic threshold
}
\end{center}

Here we report the technical details of the results presented in the Letter. The Supplementary Material is organized as follows
\begin{enumerate}
\item In Section \ref{Luttinger} we briefly discuss the moving defect in the Luttinger Liquid's approximation, showing the incapability of the latter of describing any LQSS in the present context.
\item In Section \ref{LSsec} we apply the general framework of the Lippmann-Schwinger equation to our scattering problem. In particular, in Section \ref{scatsumsec} we derive the sum rules obeyed by the scattering amplitudes and in Section \ref{planewavesec} we derive the LQSS.
\item In Section \ref{deltasec} we exactly compute the scattering data for a $\delta-$like defect.
\item In Section \ref{semiclassicsec} we present a detailed explanation of the semiclassical approximation.
\end{enumerate}

\section{The Luttinger Liquid approximation}
\label{Luttinger}

The Luttinger Liquid approach \ocite{LL1,LL2} provides a well established approximated description of the low energy sectors of several gapless one dimensional models, among those the hopping fermions system at hand. However, it is not limited to free models and it widely succeeds in exploring non-perturbative effects in interacting models. It is then natural to investigate the moving defect in this approximation scheme, aiming at identifying the universal features. 
However, when applied to the moving impurity, the Luttinger Liquid is unable to capture the out-of-equilibrium steady state that is observed on the lattice.
The Luttinger Liquid maps the low energy sector of interacting fermions in a free massless bosonic field with Lagrangian
\be
\mathcal{L}_\text{LL}=\int dx\,\frac{1}{K v_\text{s}} \partial_\mu\phi\partial^\mu\phi\, .
\ee
Where $K$ is the adimensional Luttinger parameter, whose value depends on the interaction ($K=1$ for free fermions), $v_s$ is the sound velocity that it appears also in the covariant derivative $\partial_\mu=(\partial_t,v_\text{s}\partial_x)$. Correlation values of the fields are recovered from the \emph{vertex operators} and the descendants of the field $\phi$. For example, it holds the correspondence
\be
c^\dagger_j c_j\simeq n_0-\frac{1}{\pi}\partial_x\phi\, ,
\ee
where $n_0$ is the background constant fermionic density. In this perspective, the moving defect in the Luttinger Liquid approach would be described by a Lagrangian
\be
\mathcal{L}_\text{LL}=\int dx\,\frac{1}{K v_\text{s}} \partial_\mu\phi\partial^\mu\phi+ \frac{1}{\pi}V(x-vt)\partial_x\phi\, .
\ee
The Luttinger Liquid approximation holds as long as we remain in the low energy sector, i.e. for weak and smooth defects. This simple Luttinger description is further enriched in literature through the so-called depletion model [\ocite{imp1,imp2,imp4,imp6,imp7}], where an impurity of finite mass is phenomenologically coupled to a Luttinger Liquid mimicking the low energy excitations of the surrounding environment, but such an analysis goes beyond our current purposes.
We discuss now the solution of the out-of-equilibrium protocol: from the quantum equation of motion
\be
\partial_\mu\partial^\mu\phi=-\frac{K v_s}{\pi} \partial_x V(x-vt)\, 
\ee
we immediately see that the potential term can be absorbed through a shift of the bosonic field (if $v\ne v_s$)
\be
\phi(x,t)=\tilde{\phi}(x,t)+J(x-vt)\, 
\ee
where
\be
J(x)=-\int^x dx'\,\frac{K v_s V(x')}{\pi(v^2-v_s^2)} \, .
\ee
Above, the integration constant of the indefinite integral will not affect any observable and can be left unspecified. The field $\tilde{\phi}$ evolves with the homogeneous equation of motion $\partial_\mu\partial^\mu\tilde{\phi}=0$ and thus it can be split in left/right movers and the solution easily follows using that, at time $t=0$, the density is homogeneous $\langle\partial_x \phi\rangle=0$. In particular, we report the $t>0$ evolution of the density fluctuations, being similar conclusions valid for other observables
\be
\langle\partial_x\phi(x,t)\rangle=\frac{K v_s V(x+v_st)}{2\pi(v^2-v_s^2)}\Bigg(1-\frac{v}{v_s}\Bigg)+\frac{K v_s V(x-v_st)}{2\pi(v^2-v_s^2)}\Bigg(1+\frac{v}{v_s}\Bigg)-\frac{K v_s V(x-vt)}{\pi(v^2-v_s^2)}\, .
\ee
We can recognize three contributions. Two of them spread at sounds velocity $v_s$ from the initial position of the defect: these are simply the signal of having activated the defect that spreads in the homogeneous system. The third contribution instead moves with the defect, thus at velocity $v$. For any velocity, the LQSS is absent and not peculiar behavior emerges passing from a subsonic to a supersonic regime. The Luttinger Liquid result is consistent with our scattering theory, in fact an incoming particle at wavevector $k$ can scatter in an outgoing particle at vector $q$ only if it respects the energy conservation
\be
-vk-\cos(k)=-vq -\cos(q)\, .
\ee
In the Luttinger Liquid paradigm only small wavevectors around the edges of the Fermi sea are considered, thus if we linearize around the right edge (the left case is analogous) $k\to k_F+\delta k$ and $q\to k_F+\delta q$ we obtain
\be
-v\delta k-v_s\delta k=-v\delta q-v_s\delta q\, ,
\ee
that has always the unique solution (if $v\ne -v_s$) $\delta k=\delta q$. This means that the particles do not scatter and are purely transmitted by the defect without producing an LQSS. 
 As we observed in the main text, this conclusion is only true for a supersonic defect; indeed, the linearization of the dispersion relation at the basis of the LL derivation 
does not capture the several scattering channels emerging in the subsonic case.

\section{Eigenstates in the asymptotic region and Lippmann-Schwinger equation}
\label{LSsec}

The modes of the continuous Hamiltonian (Eq. (3) in the main text) in the defect reference frame are solutions of the time-independent version of 
the Schr\"oedinger equation
\be
E\psi(x)=iv\partial_x\psi(x)-\frac{\psi(x+1)+\psi(x-1)}{2}+V(x)\psi(x)\, .\label{sch}
\ee
Assuming a localized potential around $x=0$, we can use the scattering theory to describe the eigenfunctions. In particular, the latter can be labeled in terms of the asymptotic in-states (plane waves) and must satisfy the Lippman-Schwinger equation [\ocite{sakurai}]
\be
|\psi_k\rangle=|k\rangle+\frac{1}{E(k)-T+i0^+}V|\psi_k\rangle\, ,\label{lipSM}
\ee
where we adopted a bra/ket notation.  Above, $\langle x|k\rangle=e^{ikx}/\sqrt{2\pi}$, $E(k)=-\cos(k)-vk$ and we split the Hamiltonian in the kinetic and potential term
\be
\langle x|T|\psi_k \rangle=iv\partial_x\psi_k(x)-\frac{\psi_k(x+1)+\psi_k(x-1)}{2}\, ,\hspace{2pc} \langle x|V|\psi_k\rangle=V(x)\psi_k(x)\, .
\ee
The kinetic term is clearly diagonal in the momentum basis $\langle k|T|q\rangle=\delta(k-q)E(k)$: inserting a decomposition of the identity in the momentum basis in Eq. (\ref{lipSM}) and contracting it with $\langle x|$, we immediately recover the Lippmann-Schwinger equation in the coordinate space
\be
\psi_k(x)=\frac{e^{ikx}}{\sqrt{2\pi}}+\int dx' \,\mathcal{K}_k(x-x')V(x')\psi_k(x')\, ,\label{coordsw}
\ee
where the kernel $\mathcal{K}_k(x)$ is defined as
\be
\mathcal{K}_k(x)=\int \frac{dq}{2\pi}\frac{e^{iqx}}{E(k)-E(q)+i0^+}\,  .\label{int_kernel}
\ee
Aiming to describe a scattering theory, we are interested in the behavior of $\psi_k(x)$ in 
the asymptotic region $|x|\to \infty$. 
In this limit the integral in \eqref{int_kernel} can be closed in the upper (lower) complex half-plane for $x>0$ ($x<0$).
Then, all the poles with a non-vanishing imaginary part only give exponential corrections to the contribution of 
singularities on the real axis. This leads to
\be
\mathcal{K}_k(x)=\sum_{k_n}\frac{\theta(v(k_n) x)}{i|v(k_n)|}e^{ik_n x} +...\hspace{4pc} |x|\to \infty\label{lipp_kern}\, ,
\ee
where $\theta$ is the Heaviside Theta function and $v(k)$ is the classical velocity $v(k)=\sin k-v$. The scattered wavevectors $k_n$ are all and only 
the solutions of $E(k)=E(k_n)$, i.e. the energy conservation condition. 
The wavefunction in the asymptotic region is thus immediately derived
\be
\psi_k(x)=\theta(-xv(k))\frac{e^{ikx}}{\sqrt{2\pi}}+\sum_{k_n} S_{k\to k_n}\theta(x v(k_n)) \frac{e^{ik_n x}}{\sqrt{2\pi}}\, .\label{asymptSM}
\ee
Where
\be
S_{k\to k_n}=\begin{cases}\frac{\sqrt{2\pi}}{i|v(k_n)|}\int dk\,e^{-ik_n x'}V(x')\psi_k(x')=\frac{2\pi}{i|v(k_n)|}\langle k_n|V|\psi_k\rangle \hspace{2pc} &k_n\ne k \\
\, &   \\
1+\frac{\sqrt{2\pi}}{i|v(k)|}\int dk\,  e^{-ik x'}V(x')\psi_k(x')=1+\frac{2\pi}{i|v(k)|}\langle k|V|\psi_k\rangle \hspace{2pc} &k_n= k
\end{cases}
\ee
Of course the above expressions are only an \emph{implicit solution} for the scattering amplitudes that cannot be determined without the exact knowledge of $\psi_k$. This issue will be consider in Section \ref{deltasec} and Section \ref{semiclassicsec}, thereafter we instead derive some conclusions valid for generic scattering amplitudes.

\subsection{Sum rules for the scattering amplitudes}
\label{scatsumsec}

The scattering amplitudes that appear in (\ref{asymptSM}) must obey proper sum rules induced by the unitarity of the Lippmann-Schwinger equation. Here we provide a derivation of these sum rules for the scattering problem at hand.
The Lippmann-Schwinger equation is equivalent to unitarily evolve the asymptotic in-state with an adiabatic turning on of the defect [\ocite{sakurai}], thus it preserves the scalar products and we must have $\braket{\psi_k|\psi_q}=\braket{k|q}=\delta(k-q)$.
Computing the scalar product $\braket{\psi_k|\psi_q}$, the singular behavior is due only to the wavefunction in the asymptotic region: thus we can compute $\braket{\psi_k|\psi_q}$ using directly (\ref{asymptSM}).
We now impose the conservation of the scalar product
\begin{eqnarray}
\nonumber&&\delta(k-q)=\int dx\, \psi^*_k(x)\psi_q(x)=\left[\sum_{n,m}S^*_{q\to q_n}S_{k\to k_m}\theta(v(k_m)v(q_n))\int \frac{dx}{2\pi}\theta(xv(k_m))e^{i(k_m-q_n)x}+\right.\\
\nonumber&& +\sum_{n}S^*_{q\to q_n}\theta(-v(k)v(q_n))\int \frac{dx}{2\pi}\theta(-xv(k))e^{i(k-q_n)x}+\sum_{n}S_{k\to k_m}\theta(-v(k_m)v(q))\int \frac{dx}{2\pi}\theta(-xv(q))e^{i(k_m-q)x}+\\
&&\left.+\theta(v(k)v(q))\int \frac{dx}{2\pi}\theta(-xv(q))e^{i(k-q)x}\right]+...\label{scalar}
\end{eqnarray}
The neglected terms are the corrections due to the replacement of the true eigenfunctions with their approximation in the asymptotic region. 
Integrating the $\theta$ functions according to
\be
\int dx\, \theta(x) e^{i\omega x}=iP\left(\frac{1}{\omega}\right)+\pi\delta(\omega)\, ,
\ee
where $P(\ldots)$ is the principal value, we obtain the sought singular terms. 
Equating the coefficients $\delta-$singular parts in the left and right-hand side of (\ref{scalar}) the following equality is obtained
\be
\sum_{q_n,k_m}S^*_{q\to q_n}S_{k\to k_m}\delta(k_m-q_n)=\delta(k-q)\, ,
\ee
where all the $q_n$ and $k_m$ are respectively functions of $k$ and $q$, given by the energy conservation $E(k_m) = E(k)$
and $E(q_n) = E(q)$. 
In order to compare the non singular pre-factors of the two members, the Jacobian $dk_m/dk$ (and $dq_n/dq$)  must be taken into account. Using the fact that $E(k)=E(k_n)$ we readily have
\be
\frac{d k_m}{d k}=\frac{d E(k_m)}{d k_m}\left(\frac{d E(k)}{d k}\right)^{-1}=\frac{v(k)}{v(k_m)}\label{jacob}\, .
\ee
With this last piece of information (and a slight change of notation in order to have a compact expression) we obtain the orthogonality condition
\be
\sum_{k_m} S^*_{k_n\to k_m}S_{k_{n'}\to k_{m}}|v(k_m)|=|v(k_n)|\delta_{n,n'}\, .\label{ortho1}
\ee
The physical interpretation of the above is especially simple when $k_n=k_{n'}$.
In this case it is nothing else than the familiar normalization for the scattering matrix that sums to unity. As 
usually the incoming particles and outgoing particles have the same velocity (a part from signs) 
the extra factors $|v(k_n)|$ do not appear. We can interpret the above as a conservation of flux: the flux of outgoing particles is given by the probability of having such a particle (i.e. the modulus square of the scattering amplitude) times its velocity.
Another sum rule can be obtained similarly: being $\ket{\psi_k}$ a complete set we must have
\be
\int dk \,\braket{q|\psi_k}\braket{\psi_k|k'}=\delta(q-k')\, .
\ee
As in the previous calculations, the singular behavior is due to the asymptotic wavefunction (\ref{asymptSM}). The derivation of the forthcoming orthogonality relations is slightly more cumbersome compared with the previous case. In this approximation the scalar product is:
\be
\braket{\psi_k|k'}=\int dx\,\psi_k^*(x) \frac{e^{ik' x}}{\sqrt{2\pi} }=\frac{1}{2\pi}\left(\frac{i}{\text{sgn}(v(k))(k-k')+i\epsilon}+\sum_{k_n} \frac{iS^*_{k\to k_n}}{-\text{sgn}(v(k_n))(k_n-k')+i\epsilon}\right)+...\, ,\label{overlap}
\ee
where the limit $\epsilon\to 0^+$ must be taken. Thus
\begin{eqnarray}
\nonumber&&\int dk\,\braket{q|\psi_k}\braket{\psi_k|k'}=\int \frac{dk}{(2\pi)^2}\left(\frac{i}{\text{sgn}(v(k))(k-k')+i\epsilon_1}+\sum_{k_n} \frac{iS^*_{k\to k_n}}{-\text{sgn}(v(k_n))(k_n-k')+i\epsilon_1}\right)\times\\
&&\left(\frac{i}{-\text{sgn}(v(k))(k-q)+i\epsilon_2}+\sum_{k_{n'}} \frac{iS_{k\to k_{n'}}}{\text{sgn}(v(k_{n'}))(k_{n'}-q)+i\epsilon_2}\right)+...\, .
\end{eqnarray}
Above we must take the limits $\epsilon_1\to 0^+$ and $\epsilon_2\to 0^+$ \emph{independently}. Expanding the brackets we obtain several terms
\begin{eqnarray}
\nonumber&&\int dk\,\braket{q|\psi_k}\braket{\psi_k|k'}=\int \frac{dk}{(2\pi)^2}\Bigg(\frac{i}{\text{sgn}(v(k))(k-k')+i\epsilon_1}\frac{i}{-\text{sgn}(v(k))(k-q)+i\epsilon_2}+\\
\nonumber&&+\sum_{k_n} \frac{iS^*_{k\to k_n}}{-\text{sgn}(v(k_n))(k_n-k')+i\epsilon_1}\frac{i}{-\text{sgn}(v(k))(k-q)+i\epsilon_2}+\\
\nonumber&&+\sum_{k_{n'}} \frac{iS_{k\to k_{n'}}}{\text{sgn}(v(k_{n'}))(k_{n'}-q)+i\epsilon_2}\frac{i}{\text{sgn}(v(k))(k-k')+i\epsilon_1}+\\
&&+\sum_{k_{n'},k_n} \frac{iS_{k\to k_{n'}}}{\text{sgn}(v(k_{n'}))(k_{n'}-q)+i\epsilon_2}\frac{iS^*_{k\to k_n}}{-\text{sgn}(v(k_n))(k_n-k')+i\epsilon_1}\Bigg)\, .\label{multilineS}
\end{eqnarray}
The singular part is associated with the singularities on real axis when $\epsilon_{1,2}\to 0^+$, however the singular terms are produced only when the two poles in each of the above terms pinch the real axis in the limit $\epsilon_{1,2}\to 0^+$. This is verified in the first and last line.
For example, consider the first integral 
\begin{eqnarray}
\nonumber&&\int \frac{dk}{(2\pi)^2}\frac{i}{\text{sgn}(v(k))(k-k')+i\epsilon_1}\frac{i}{-\text{sgn}(v(k))(k-q)+i\epsilon_2}\simeq \int \frac{dk}{(2\pi)^2}\frac{1}{(k-k')+i\text{sgn}(v(k'))\epsilon_1}\frac{1}{(k-q)-i\text{sgn}(v(q))\epsilon_2}=\\
&&\simeq \frac{2\pi i}{(2\pi)^2}\frac{-\text{sgn}(v(k'))}{(k'-q)-i\text{sgn}(v(k'))(\epsilon_1+\epsilon_2)}=\frac{1}{2}\delta(k'-q)+...
\end{eqnarray}
where we neglected non singular terms at each passage. Thus, the singular part of (\ref{multilineS}) is:
\be
\delta(k'-q')=\int dk\,\braket{q|\psi_k}\braket{\psi_k|k'}=\frac{1}{2}\delta(k'-q)+\frac{1}{2}\sum_{k_{n'},k_n}\int dk S_{k\to k_{n'}}S^*_{k\to k_n}\delta(k_{n'}-q)\delta(k_n-k')\,.
\ee
Using Eq. (\ref{jacob}) as before to compare the non singular amplitudes of the above equation, we get a second orthogonality condition
\be
\sum_{k_n} \frac{S_{k_n\to k_m}S^*_{k_n\to k_{m'}}}{|v(k_n)|}=\frac{\delta_{m,m'}}{|v(k_m)|}\label{orth_full_s_app}\, .
\ee

\subsection{Evolution of the plane wave and LQSS}
\label{planewavesec}

As we reported in the Letter, the key point in the derivation of the LQSS is computing the evolution of the plane wave $\ket{k}$, at late times and in the asymptotic region
\be
f_k(x,t)=\sqrt{2\pi}\int dq\, e^{-iE(q) t}\psi_q(x)\langle \psi_q| k\rangle\, .
\ee
Being interested in the asymptotic region, we replace $\psi_q(x)$ with (\ref{asymptSM}) at the price of neglecting 
contributions exponentially vanishing away from the defect. This leads to 
\be
f_k(x,t)=\int dq\, e^{-iE(q) t}\left(\theta(-xv(q))e^{iqx}+\sum_{q_n} S_{q\to q_n}\theta(x v(q_n)) e^{iq_n x}\right)\langle \psi_q| k\rangle\, .
\ee
Even though we do not know exactly the overlap $\langle \psi_q|k\rangle$, being interested in the large time limit we can replace $\psi_q(x)$ with its expression in the asymptotic region (\ref{asymptSM}), i.e. Eq. (\ref{overlap}). In fact, only the singularities of the integrand contribute to the late time behavior and any non singular 
contribution decays because of the fast oscillating phases $e^{iq_n x-iE(q)t}$. At fixed ray $\zeta=x/t$, the neglected contribution 
will vanish as $\sim t^{-1/2}$ as one sees via the saddle point approximation (a slower decay $\sim t^{-1/3}$ is obtained 
if $\zeta$ is at the edges of the lightcone).
Thus in this approximation we have
\begin{eqnarray}
\nonumber&&f_k(x,t)=\int \frac{dq}{2\pi}\, e^{-iE(q) t}\left(\theta(-xv(q))e^{iqx}+\sum_{q_n} S_{q\to q_n}\theta(x v(q_n)) e^{iq_n x}\right)\times \\
&&\left(\frac{i}{\text{sgn}(v(q))(q-k)+i\epsilon}+\sum_{q_{n'}} \frac{iS^*_{q\to q_{n'}}}{-\text{sgn}(v(q_{n'}))(q_{n'}-k)+i\epsilon}\right)\, .\label{f1}
\end{eqnarray}

Again, only the singular contributions survive at late times. Using the identity (in the distribution sense)
\be
\lim_{t\to\infty,\, \zeta \text{ fixed}} \frac{ie^{-iE(q)t +iq_n x }}{-\text{sgn}(v(q_{n'}))(q_{n'}-k)+i\epsilon}=2\pi e^{-iE(k)t+ik_{n} x}  \theta\left[\text{sgn}(v(k_{n}))\left( \zeta-v(k_n)\right)\right]\delta(q_{n'}-k)\label{dis_lim}\, ,
\ee
we can compute the asymptotic evolved plane waves. Again, correction to Eq. (\ref{dis_lim}) can be estimated through the saddle point method and can be proved to vanish at least as $\sim t^{-1/3}$.
Using (\ref{dis_lim}) in (\ref{f1}) and (\ref{jacob}) to carry on the explicit integration over the $\delta-$functions we readily get
\begin{eqnarray}
\nonumber&&e^{iE(k) t}f_{k}(x,t)=\theta(-\zeta v(k)) e^{ikx} +\sum_{k_n} e^{ik_n x}  \theta(\zeta v(k_n))\theta\left[\left(|v(k_n)|-|\zeta|\right)\right]S_{k\to k'_n}+\\
&&+\sum_{q,q_n,q_{n'}} \left[e^{iq_{n} x} \theta(\zeta v(q_n)) \theta\left(|\zeta|- |v(q_n)|\right)S_{q\to q_n}S^*_{q\to q_{n'}}\left|\frac{v(q_{n'})}{v(q)}\right|\right]_{q_{n'}=k} \, .
\end{eqnarray}

The summation of the last line is greatly simplified by the orthogonality condition (\ref{orth_full_s_app}), leading to the final result we were looking for
\be
\lim_{t\to\infty}e^{iE(k)t}f_{k}(x,t)= e^{ik x}\theta(-\zeta v(k))+e^{ik x}\theta(\zeta v(k)) \theta\left( |\zeta|-|v(k)|\right)+\sum_{k_n} e^{ik_n x}  \theta(\zeta v(k_n))\theta\left(|v(k_n)|-|\zeta|\right)S_{k\to k_n}\, .\label{evplane}
\ee

With this last result, we can analyze the two point correlator 
\be
\langle \tilde{c}^\dagger_x\tilde{c}_y\rangle_t=\int_{-k_f}^{k_f}\frac{dk}{2\pi} f^*_k(x,t)f_k(y,t)\,\label{originalcorr}
\ee
in the large time limit. Inserting (\ref{evplane}) in the above we get a sum of several oscillating phases: being interested in the LQSS scaling limit we can rely on the further assumption of large distances from the defect and thus discard several contributions, finally leading to the desired 
expression
\be
\langle \tilde{c}^\dagger_x\tilde{c}_y\rangle_t=\int_{-k_f}^{k_f}\frac{dk}{2\pi} e^{ik(y-x)}\left[1-\theta(\zeta v(k))\theta(|v(k)|-|\zeta|)\right]+\sum_{k_n}|S_{k\to k_n}|^2\theta(\zeta v(k_n))\theta(|v(k_n)|-|\zeta|)e^{ik_n(y-x)}\, .\label{lqss}
\ee
Discarding the oscillating corrections to (\ref{lqss}) requires some additional care. Despite being vanishing corrections, simple saddle point arguments predict a slow algebraic $(\sim 1/\sqrt{|x|}\simeq 1/\sqrt{|y|})$ decay that can affect the LQSS on a practical purpose.
In particular, such oscillations are relevant for the NESS description, when only time is sent to infinity while the distance from the defect is kept finite. Persistent oscillations are found in the case of a $\delta-$like defect, that we are going to analyze.

\section{An exact result: the $\delta-$like defect}
\label{deltasec}
In this section we determine the \emph{exact} eigenfunctions of the time-independent Schr\"oedinger equation (\ref{sch}) in the case of a $\delta-$like defect, i.e. $V(x)=c\delta(x)$. Obviously, the exact eigenfunctions permit an exact determination of the scattering amplitudes $S_{k\to k_n}$ and thus of the LQSS generated by such a defect.
Attempting to solve directly the Schr\"oedinger equation is complicated because of the presence of the hopping term, on the other hand the Lippmann-Schwinger equation leads to some ambiguities for the limiting case $V(x)=c\delta(x)$: the best way to solve the problem is combining the two approaches.
Consider the Lippman-Swinger equation in the coordinate space (\ref{coordsw}), specializing to a $\delta-$like potential we have
\be
\psi_k(x)=\frac{e^{ikx}+\mathcal{K}_k(x)\alpha_k}{\sqrt{2\pi}}\label{delta_wave}\, ,
\ee
where $\alpha_k$ is a constant formally defined as 
\be
\alpha_k=c\sqrt{2\pi}\int dx' \delta(x')\psi_k(x')\, .
\ee
We could be tempted of explicitly performing the integral by the $\delta$-function. 
However, $\psi_k$ is not continuous in $x=0$ (this simply because $\mathcal{K}_k(x)$ is discontinuous in $x=0$) and therefore
\be
\int dx' \,\delta(x')\psi_k(x') \neq \psi_k(0)\hspace{2pc} \label{deltaint}\, . 
\ee
The arbitrariness in the regularization of (\ref{deltaint}) is resolved looking directly at the Schr\"oedinger equation (\ref{sch}), that we report below for convenience 
\be
E(k)\psi_k(x)=iv\partial_x\psi_k(x)-\frac{\psi_k(x+1)+\psi_k(x-1)}{2}+c\delta(x)\psi(x)\, .
\ee
We consider now what happens at the singularity in $x=0$, matching the singular terms in the above: obviously $\delta(x) \psi_k(x)$ is singular. Then, $\psi_k$ cannot be singular on its own and this leaves out only $i v\partial_x\psi_k$ as a possible singular term. Thus, retaining in the above only the singular terms, we formally get
\be
iv\partial_x\psi_k+c\delta(x)\psi_k(x)=0\hspace{1pc}\Rightarrow\hspace{1pc} \psi_k(x)=e^{i\frac{c}{v}\int^x_{x_0}dx'\,\delta(x')}\psi_k(x_0)\, .
\ee
Thus, the discontinuity of the wavefunction in $x=0$ is determined as
\be
\psi_k(0^+)=e^{i\frac{c}{v}}\psi_k(0^-)\, .
\ee
We can supplement the explicit form of the wavefunction (\ref{delta_wave}) with this last information and determine the constant $\alpha_k$ 
\be
\begin{cases}\psi_k(0^+)=\frac{1}{\sqrt{2\pi}}\left(1+\mathcal{K}_k(0^+)\alpha_k \right)\\ \, \\
\psi_k(0^-)=\frac{1}{\sqrt{2\pi}}\left(1+\mathcal{K}_k(0^-)\alpha_k \right)\\ \, \\
\psi_k(0^+)=e^{i\frac{c}{v}}\psi_k(0^-)
\end{cases} \hspace{1pc}\Longrightarrow\hspace{1pc}\alpha_k=\frac{1-e^{i\frac{c}{v}}}{e^{i\frac{c}{v}}\mathcal{K}_k(0^-)-\mathcal{K}_k(0^+)}\label{alphak}\, .
\ee

\begin{figure*}[t!]
\includegraphics[width=0.45\textwidth]{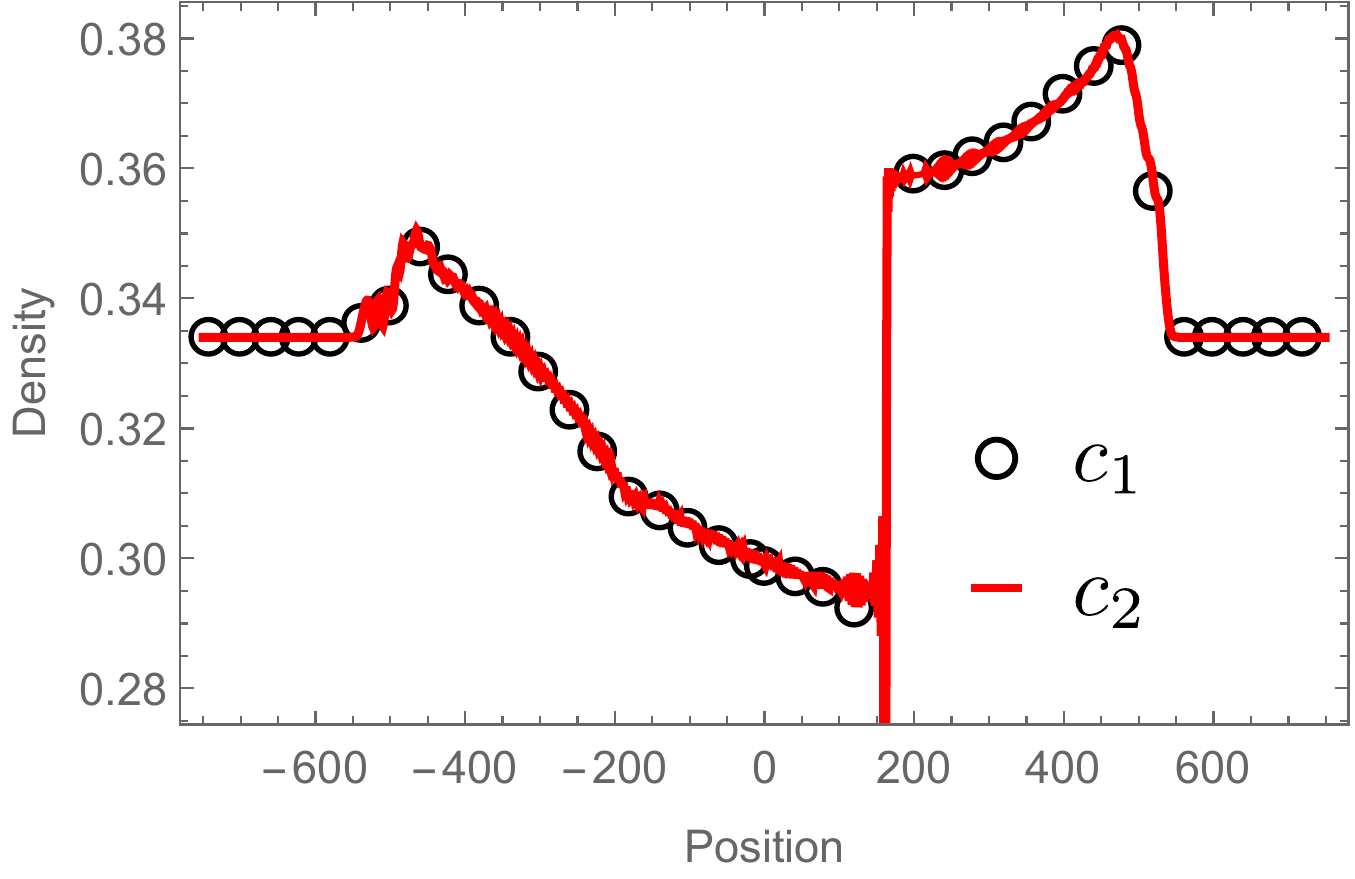}
\caption{\label{two_values}The figure displays the numeric fermionic density $\langle d_j^\dag d_j \rangle$ as a function of the position at a fixed time ($t=530$) for two $\delta-$like defects $V(x)=c\delta(x)$, moving at the same velocity $v=0.3$, but with different strengths, respectively $c_1=0.5$ and $c_2=c_1+2\pi v$. 
Oscillations are smeared out by averaging on neighboring sites (see also FIG. \ref{scat} and the related discussion).
}
\end{figure*}

The final step is computing $\mathcal{K}_k(0^\pm)$ through a direct analysis of the defining integral (\ref{int_kernel}).
However, before doing so, it is worth noticing a simple fact with interesting physical consequences: $\alpha_k$, as it is clear from eq. (\ref{alphak}), is a periodic function of $c/v$ with period $2\pi$. Therefore there exist different values of the coupling sharing exactly the same LQSS, in particular for $c/v=2\pi n$ with $n\in \mathbb{Z}$ the LQSS is never produced, even in the case of a subsonic defect (see FIG. \ref{two_values}). The periodicity in the coupling does not concern only the LQSS, but the entire time evolution, since such periodicity is fulfilled by the exact eigenvectors. In particular, for $c/v=2\pi n$ the eigenvectors simply become the plane waves, i.e. the eigenvectors of the Hamiltonian in absence of the defect: in this case neither LQSS or any transient is produced and the system is completely unaware of the traveling defect.
We now proceed in computing $\mathcal{K}_k(0^\pm)$.
The asymptotic expansion (\ref{lipp_kern}) has been extracted from the poles along the real axis in the integral (\ref{int_kernel}). Adding and subtracting the poles we can thus write
\be
\mathcal{K}_k(x)=\sum_{k_n}\frac{\theta(v(k_n) x)}{i|v(k_n)|}e^{ik_n x}+\int \frac{dq}{2\pi}\left(\frac{e^{iqx}}{E(k)-E(q)+i\epsilon}-\sum_{k_n}\frac{e^{iqx}}{-v(k_n)(q-k_n)+i\epsilon}\right)\, .
\ee
Now, the term in brackets is no longer singular when $\epsilon\to 0$, thus we can directly take the limit. Consider now the integral
\be
\tau(k,x)=\int \frac{dq}{2\pi}\left(\frac{e^{iqx}}{E(k)-E(q)}+\sum_{k_n}\frac{e^{iqx}}{v(k_n)(q-k_n)}\right)\, .
\ee
$\tau$ has a continuous and a discontinuous contribution when $x\to 0^{\pm}$. In order to extract it, we split the integral into two regions with a cut off $\Delta$
\be
\tau(k,x)=\int_{|q|<\Delta} \frac{dq}{2\pi}\left(\frac{e^{iqx}}{E(k)-E(q)}+\sum_{k_n}\frac{e^{iqx}}{v(k_n)(q-k_n)}\right)+\int_{|q|>\Delta} \frac{dq}{2\pi}\left(\frac{e^{iqx}}{E(k)-E(q)}+\sum_{k_n}\frac{e^{iqx}}{v(k_n)(q-k_n)}\right)\label{tau}\, .
\ee
The first integral is obviously continuous when $x\to 0$, since the integration domain is finite and there are no singularities. 
For the second term, by choosing $\Delta$ to be fixed but large, we can replace $E(k)-E(q)\simeq -v q$.
We have therefore 
\be
\int_{|q|>\Delta}\frac{dq}{2\pi} \frac{e^{i q x}}{q}=2i\int_{\Delta}^\infty\frac{dq}{2\pi}\frac{\sin(qx)}{q}=\text{sgn}(x)2i \int_{\Delta |x|}^\infty\frac{dp}{2\pi}\frac{\sin(p)}{p}\xrightarrow{x\to 0^{\pm}}=\pm \frac{i}{2}\, .
\ee
Thus, the limit $x\to 0^{\pm}$ of $\tau$ is
\be
\tau(k,0^{\pm})=\lim_{\Delta\to\infty}\left[\mathcal{P}\int_{|q|<\Delta} \frac{dq}{2\pi}\frac{1}{E(k)-E(q)}\right]\mp\frac{i}{2}\left(\frac{1}{v}-\sum_{k_n}\frac{1}{v(k_n)}\right)\, .
\ee
Above, it must be used the Principal Value prescription in the integration of singular points.
Coming back to $\mathcal{K}_k$, we finally have
\be
\mathcal{K}(k,0^\pm)=\pm\frac{1}{2iv}+\sum_{k_n}\frac{1}{2i|v(k_{n})|}+\mathcal{P}\int_{-\infty}^\infty \frac{dq}{2\pi}\frac{1}{E(k)-E(q)}\, ,
\ee
that leads to the following expression for $\alpha_k$ 
\be
\alpha_k=-\Bigg[\sum_{k_n}\frac{1}{2i|v(k_{n})|}-\frac{1}{2v}\cot\left(\frac{c}{2v}\right)+\mathcal{P}\int_{-\infty}^\infty \frac{dq}{2\pi}\frac{1}{E(k)-E(q)}\Bigg]^{-1}\, .
\ee

\begin{figure*}[t!]
\includegraphics[width=0.45\textwidth]{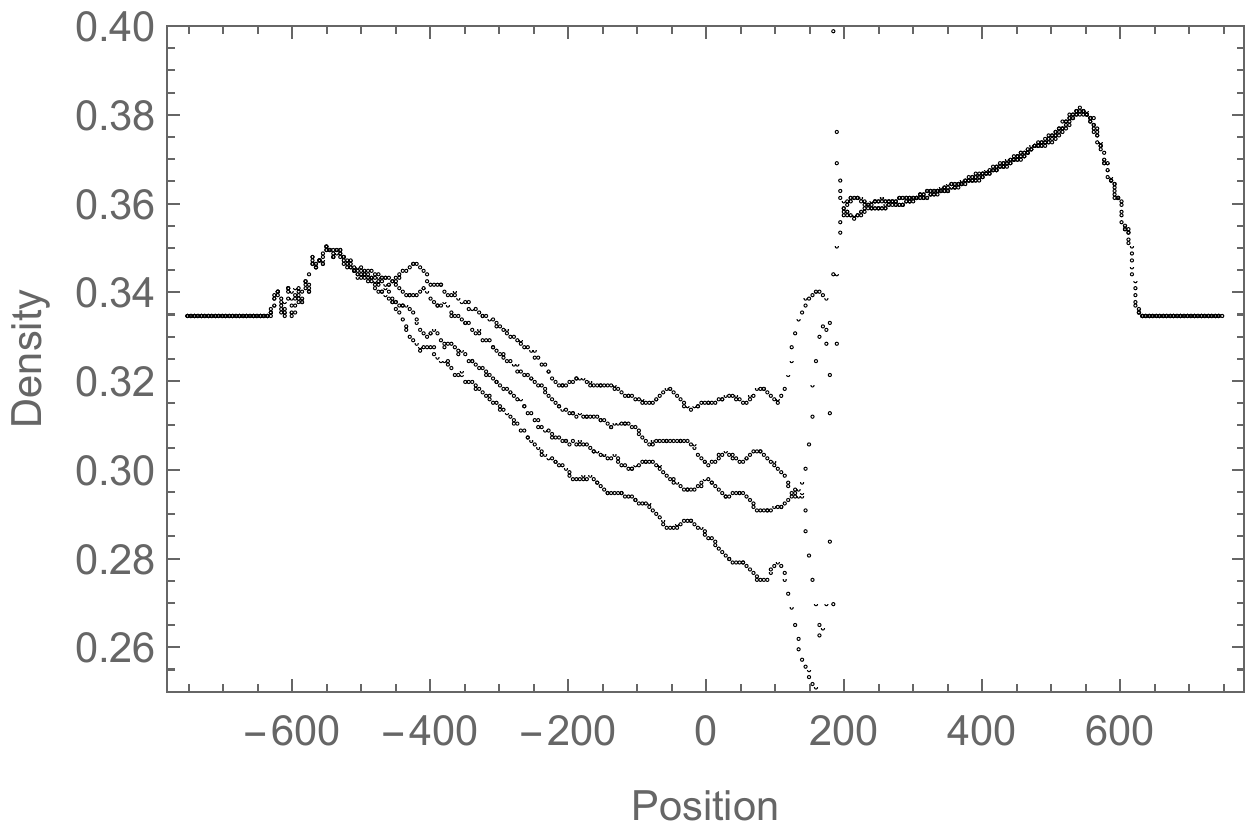}
\includegraphics[width=0.45\textwidth]{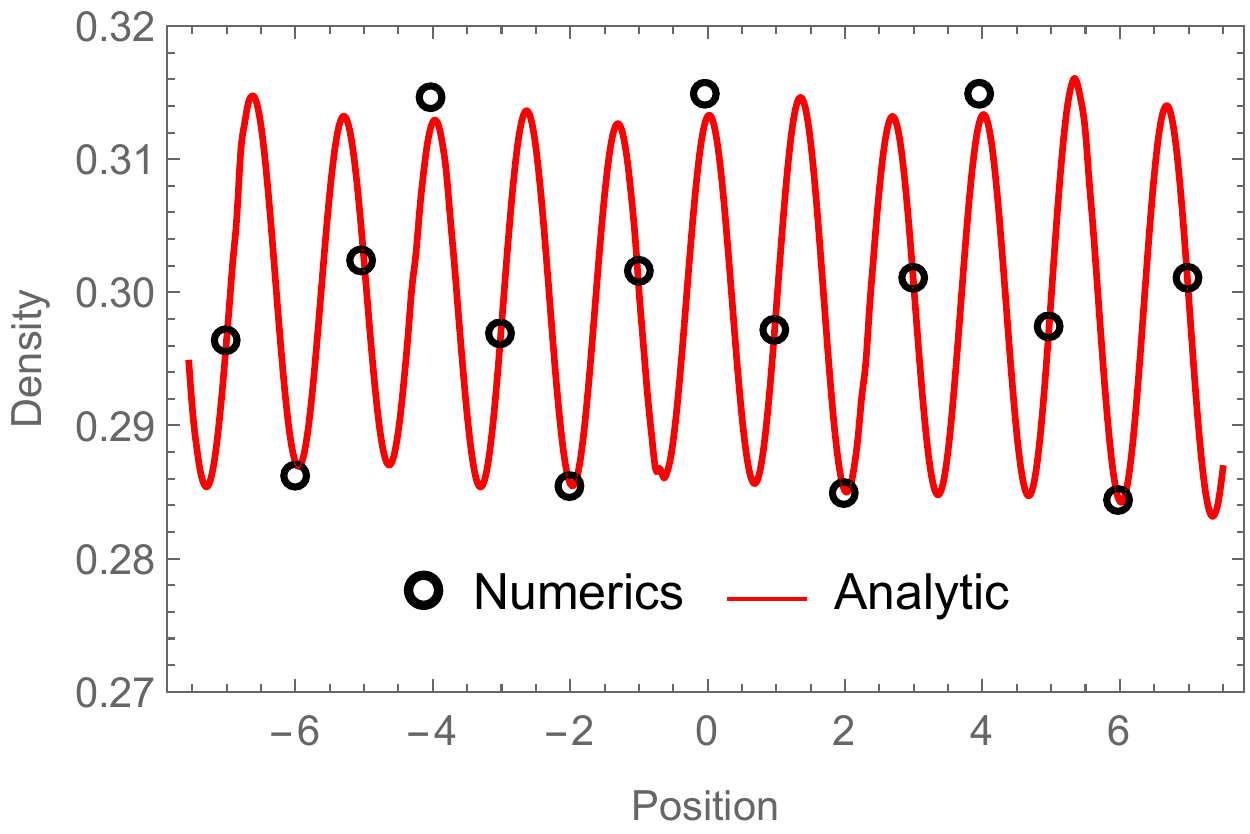}
\caption{\label{osc}
Density profile generated by a delta-like potential $V(x)=c\delta(x)$ with $c=0.5$, $v=0.3$ and $k_f=\pi/3$ at time $t=620$. Left panel: the numerical result of the whole density profile is displayed, showing slow vanishing oscillations on the left of the defect (placed where the discontinuity occurs). Right panel: the density profile is zoomed in a few-sites region around zero and the numerics is compared with the analytic prediction.}
\end{figure*}

This explicit expression permits an \emph{exact} determination of the wavefunction $\psi_k$ (\ref{delta_wave}) and therefore of the scattering amplitudes, leading to the result (8) in the main text.
In the Letter we compared the analytic LQSS with the numerics for a $\delta-$like defect in Fig. 2 of the main text. 
In that figure, we focused on the $t\to \infty$ limit which leads to the LQSS. Therefore, 
slow vanishing oscillations were smeared out with (local) spatial averages. 
If no averaging procedure is performed, the density profile looks like Fig. \ref{osc} (left panel): at fixed coordinate $x$,
the oscillations do not display any damping in time, but decay (though slowly) away from the defect, thus disappearing in the LQSS limit. \\
Nevertheless, recalling that the asymptotic wavefunction (\ref{asymptSM}) is exponentially accurate away from the defect, 
such oscillations can be captured by the scattering theory. Indeed, inserting 
the full expression of \eqref{evplane} in \eqref{originalcorr}, one obtains oscillating contribution to \eqref{lqss},
which are in agreement with the numerical simulations (right panel of Fig.~\ref{osc}).
\begin{figure*}[b!]
\includegraphics[width=0.45\textwidth]{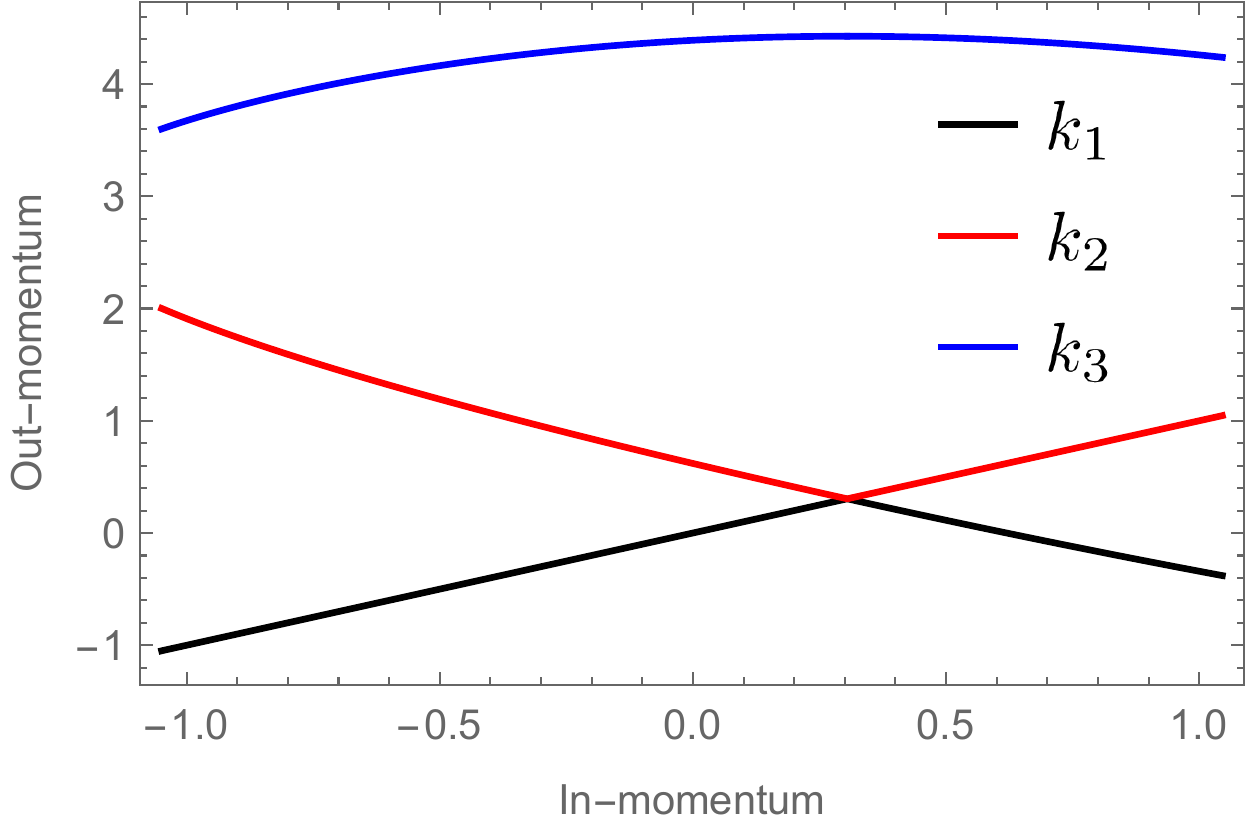}
\includegraphics[width=0.45\textwidth]{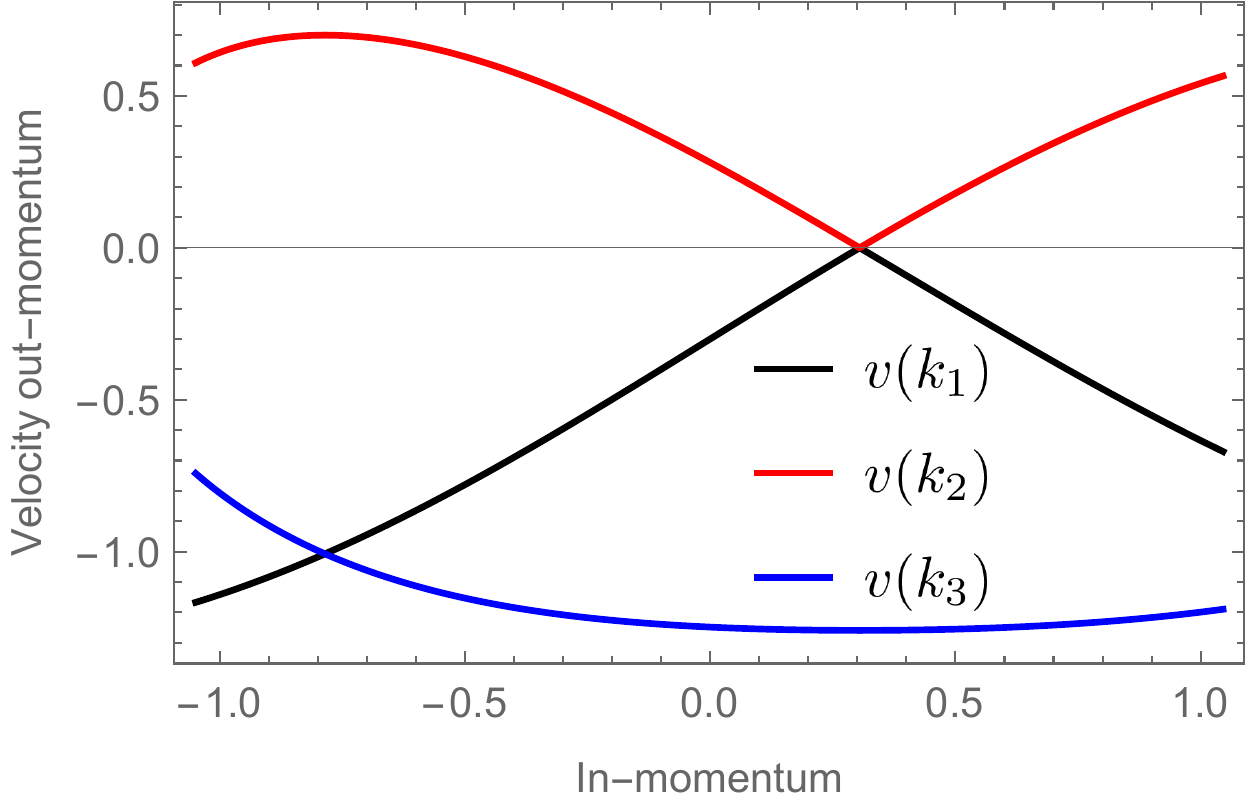}
\caption{\label{scat} Left panels: scattering channels for a moving defect at $v=0.3$ and the incoming momentum in the Fermi sea with $k_f=\pi/3$. Right panel, the velocities associated with the scattering channels. 
}
\end{figure*}
It is interesting to inspect more in detail the origin of this oscillations. For the parameters chosen in the Fig.~\ref{osc}
($v = 0.3$ and $k_f = \pi/3$), each incoming wavevector $k$ in the Fermi sea has exactly three possible scattering channels $k_1, k_2, k_3$. 
These scattering channels are plotted in Fig. \ref{scat} (left panel) together with their velocities (right panel): in the two point correlator, passing from (\ref{originalcorr}) to (\ref{lqss}), the leading corrections given by the oscillating phases $\sim e^{i (k_{n}-k_{n'})x}$ are given by the saddle point approximation ruled by the second derivative of the phase $\partial_k^2(k_{n}-k_{n'})$. As it is clear from the left panel of Fig. \ref{scat}, 
$k_n$ have a small curvature as functions of $k$ and this explains the slow vanishing oscillations. In order to understand why the oscillations in Fig. \ref{osc} appear only on the left of the defect, we should consider the velocities associated with the scattering channels (right panel Fig. \ref{scat}), together with the expression for the evolved plane wave (\ref{evplane}). For what concerns the scattering channels $k_n$, the Heaviside Theta functions in (\ref{evplane}) guarantee that the plane-wave associated with $k_n$ has support \emph{only} on the right (left) of the defect if $v(k_n)>0$ ($v(k_n)<0$).
The right panel of Fig. \ref{scat} shows that for any momentum $k$ there is only one scattering channel with positive velocity, therefore in the expression for the two point correlator (\ref{originalcorr}) no oscillating phases are produced on the right of the defect. The situation is different on the left side: as one can see in Fig. \ref{scat},
there are always two channels with negative velocity causing interference in (\ref{originalcorr}).

\section{The semiclassical approximation}
\label{semiclassicsec}

Here we present further details about the semiclassical approximation derived in the main text. Under the assumption of a smooth defect, the Wigner distribution $\rho_t(k,x)$ evolves as the density distribution of classical particles subjected to the classical equation of motion
\be
\dot{x}=v(k),\hspace{2pc} \dot{k}=-\partial_x V\, ,
\ee
that are associated with the classical Hamiltonian
\be
\mathcal{H}_\text{cl}=-\cos(k)-vk+V(x)\, .
\ee 
At initial time, because of the initial state we choose, the Wigner distribution is homogeneous in $x$ and flat in $k$ within the Fermi sea, zero outside of it.
Immediately after the defect has been switched on, the Wigner function evolves accordingly to two contributions: \emph{i)} the particles initially sat on the defect support and \emph{ii)} those that initially resided 
outside the defect support.
In the perspective of determining the LQSS, the first group of particles does not contribute at all: these particles will be scattered away from the defect at different velocities, thus they will be spread at late times, 
being irrelevant for local quantities. However, especially for large defects, the spreading of these particles could require long times.
The second group of particles is what matters at late times: the constant flux of in-going particles scatters with the defect, creating a flux of out-going particles, i.e. the classical LQSS. 
In order to describe the classical scattering events, the first step is to describe the \emph{classical trajectories}: while the solution of the equation of motion can be complicated, a rather simple picture based on the classical \emph{turning points} 
(i.e. where the particles invert their motion, thus $v(k)=0$) is more easily accessed.
The trajectories can be simply constructed following the rules reported in the main text, that we quote hereafter for convenience.
\begin{figure*}[t!]
\includegraphics[width=0.8\textwidth]{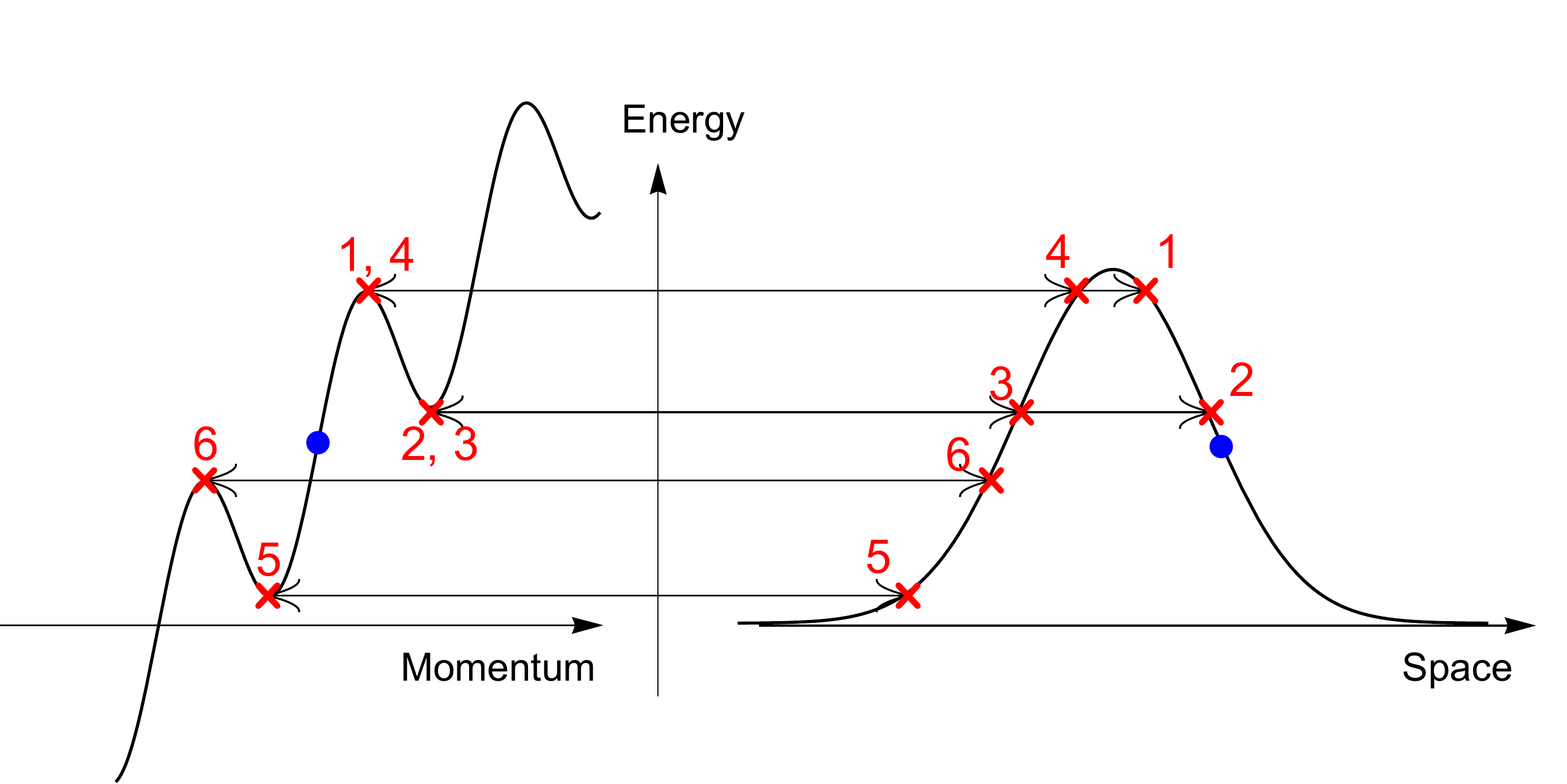}
\caption{\label{figSM1}We consider the trajectory of a classical particle initially placed in a bell-shaped potential with a given momentum. On the right the potential energy $V(x)$ is plotted as a function of position, on the left $E-\mathcal{T}(k)$, where $E$ is the initial energy. The ordinate axis of the two plots is the same to enforce the energy conservation $E-\mathcal{T}(k)=V(x)$ along the trajectory. The blue dots represent the initial condition of the particles, the red crosses are the turning points: the numeration represents the order in which the turning points are visited. The initial condition in such that $\partial_k\mathcal{T}(k)=v(k)<0$ thus the particle starts moving on the left until a turning point is reached. The first turning point to be reached is $1$, not $2$: even though $2$ appears closer than $1$ to the initial position of the particle, in the momentum space the point $2$ cannot be reached if the particle does not have crossed $1$ before. Thus, the particle travels until $1$ is reached (as long as $\partial_x V<0$, we must have $\dot{k}>0$, thus the momentum increases). After $1$ has been reached, the particle inverts its motion and the next turning point is $2$, then the particle inverts its motion again: there are no longer turning points able to stop the particles before it reaches the other side of potential (i.e. $\partial_x V>0$). Here the motion in the momentum space is reverted, thus the particle reaches again the same turning point in the momentum space, but differently placed in the space plot (point $3$). As the particle's journey proceeds, the last turning point to be visited is $6$: then the particle keeps traveling toward the left forever, its momentum being the solution of $\mathcal{T}(k)=E$ closer to the turning point $6$.
}
\end{figure*}
\begin{enumerate}
\item Until a turning point is reached, the particle maintains the same direction. When a turning point is reached, the particle inverts its motion.
\item The momentum $k$ changes continuously, moreover from $\dot{k}=-\partial_x V$ it can be seen that $k$ increases if $V$ has a negative slope and decreases otherwise.
\end{enumerate} 

\begin{figure*}[t!]
\begin{center}
\includegraphics[width=0.4\textwidth]{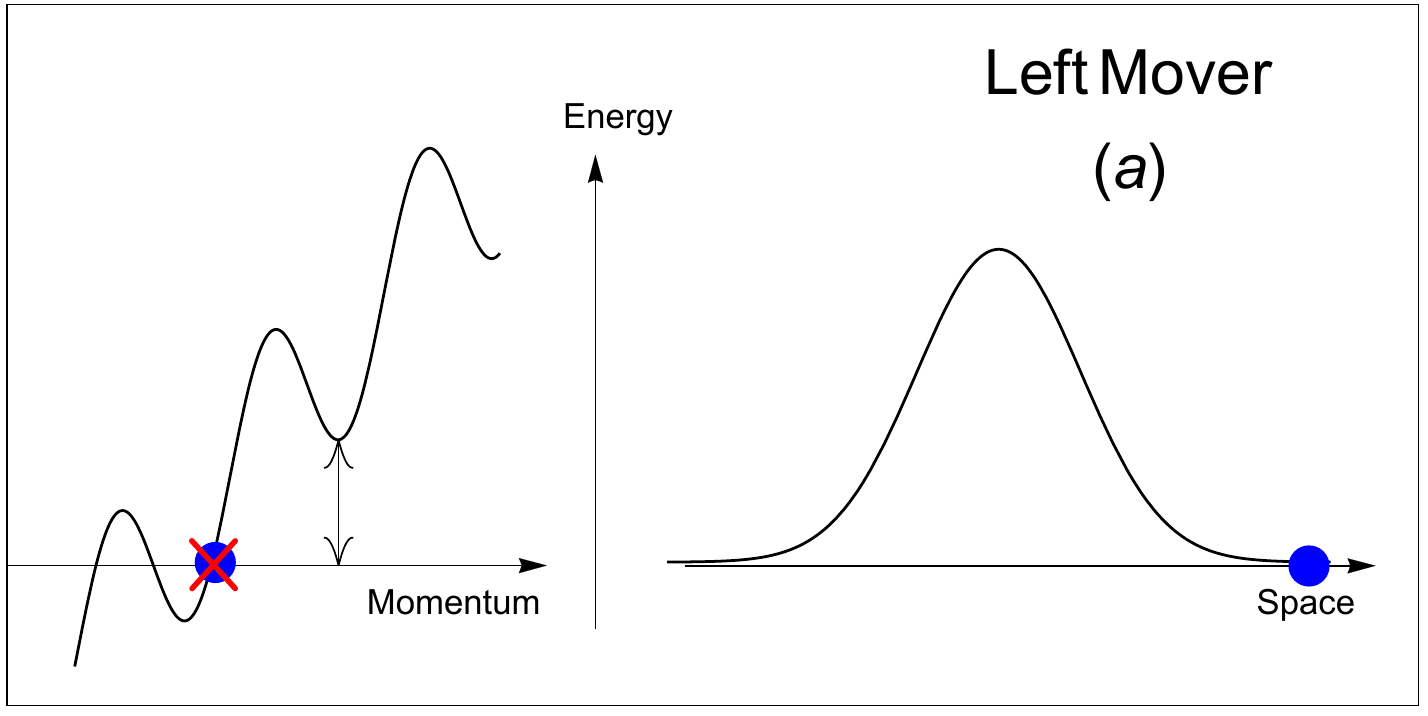}
\includegraphics[width=0.4\textwidth]{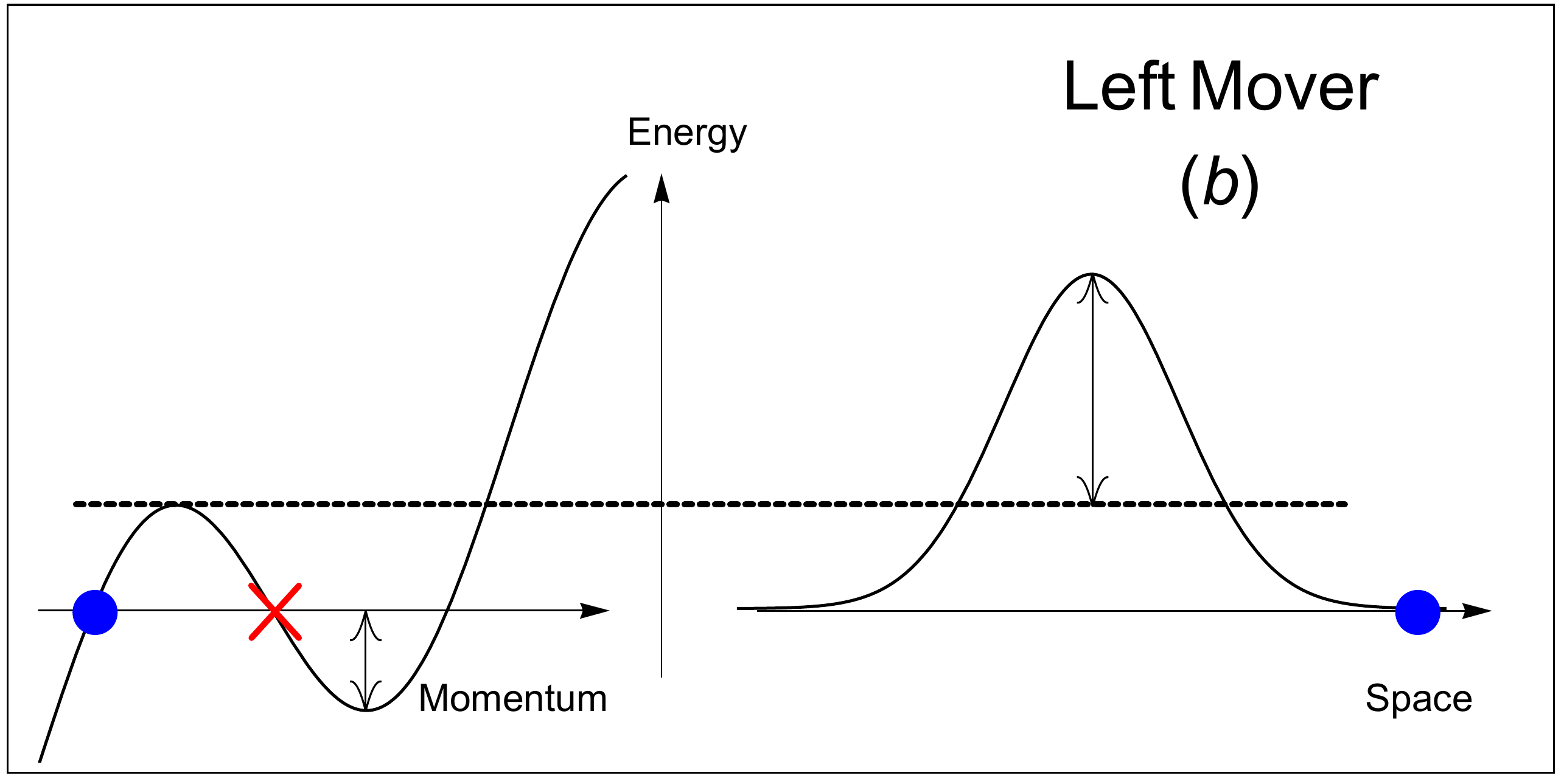}
\end{center}
\begin{center}
\includegraphics[width=0.4\textwidth]{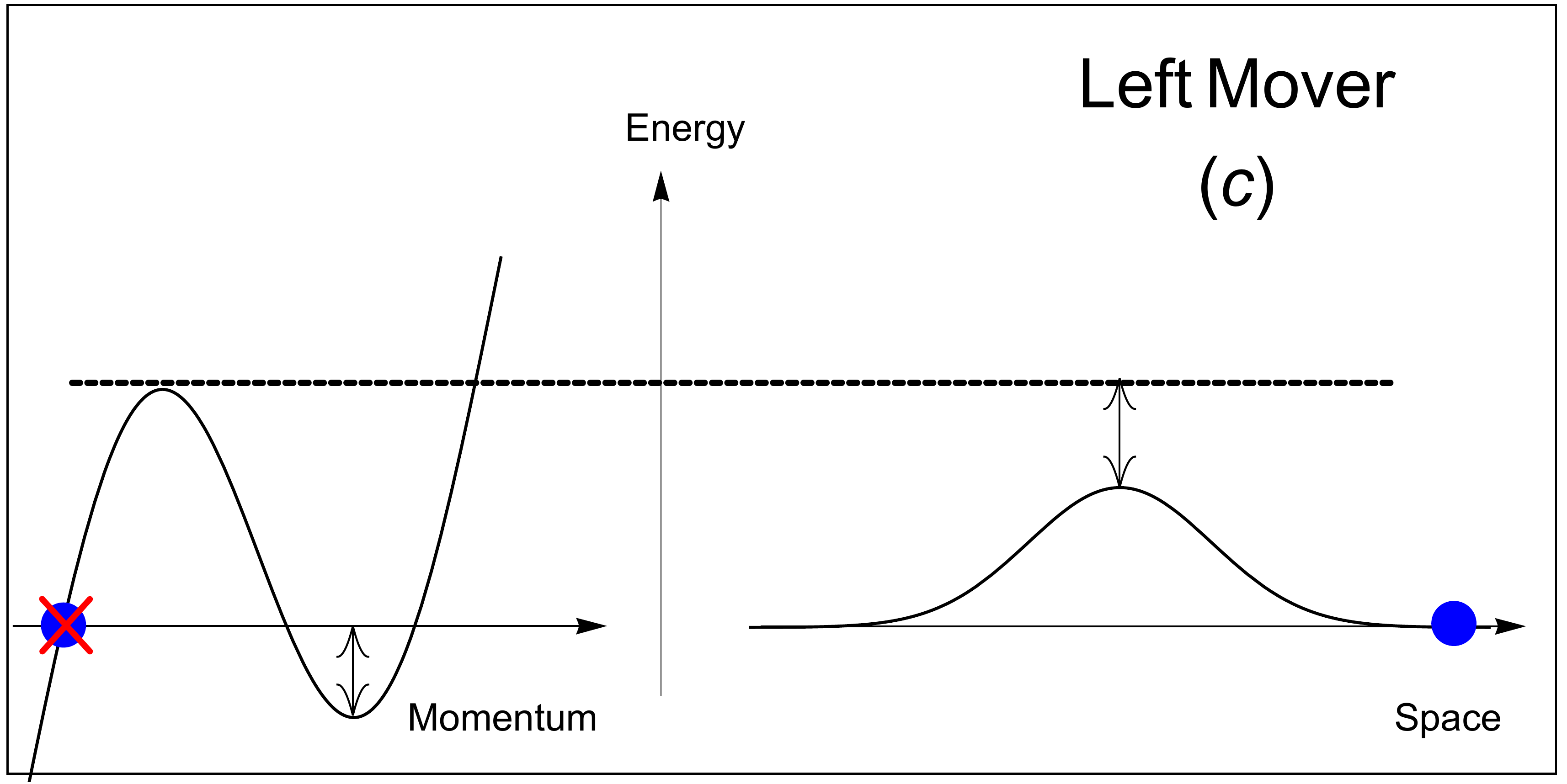}
\includegraphics[width=0.4\textwidth]{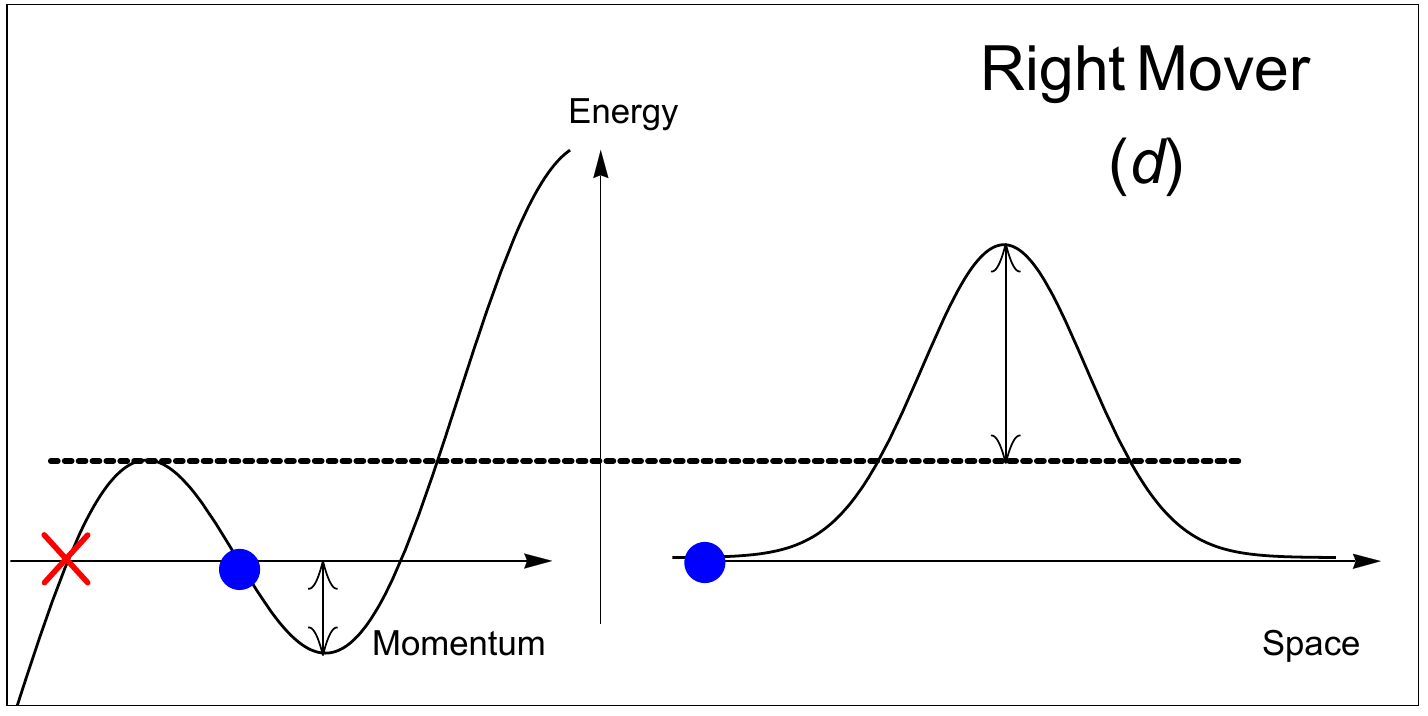}
\end{center}
\begin{center}
\includegraphics[width=0.4\textwidth]{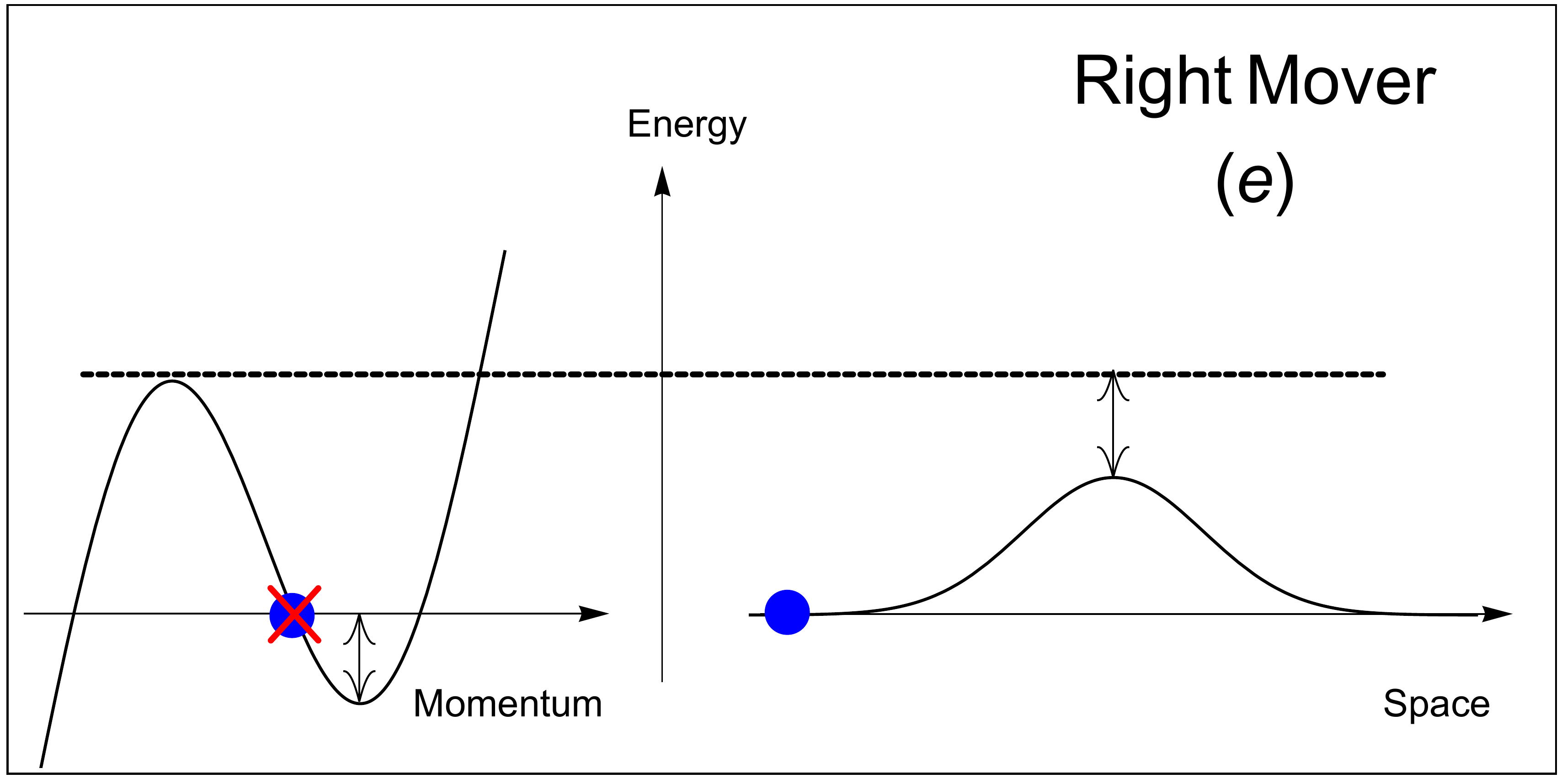}
\end{center}
\caption{\label{figSM2} In this figure the various possible scattering events on a bell-shaped repulsive potential with positive velocity are illustrated, the blue dots represent the initial position of the particle and the red crosses are the final momenta after the scattering event occurred. We start discussing the richer phenomenology of particles coming from the right, thus left movers. If the kinetic energy of the incoming particle is less than the kinetic energy of the second nearest turning point with momentum larger than the initial momentum (panel $(a)$), then the particle is transmitted independently on the potential height. If instead it is not the case, the particle is reflected if the height of the defect is larger than the difference between the kinetic energy of the nearest turning point with momentum larger than the initial one and the initial kinetic energy (panel $(b)$), if instead the height of the barrier is smaller the particle is transmitted (panel $(c)$). The destiny of the right movers depends on the height of the defect: if the height is larger than the difference in kinetic energy between the nearest turning point smaller than the initial momentum and the initial kinetic energy, then the particle is reflected, otherwise is transmitted (panels $(d)$ and $(e)$ respectively).
}
\end{figure*}

Thereafter, we discuss the determination of the trajectory of a classical particle on the bell-shaped potential of Fig. \ref{figSM1}. Along the trajectory the classical energy is conserved, so let $\mathcal{T}(k)$ be the kinetic energy $\mathcal{T}(k)=-\cos(k)-v k$, we must have $\mathcal{T}(k(t))+V(x(t))=E$. Actually, it is convenient to think about the energy conservation as $E-\mathcal{T}(k(t))=V(x(t))$ and plot, on the same energy axis, the two members (Fig. \ref{figSM1}): in this picture, the turning points are nothing else than the stationary points of the kinetic part. The particle will be initially placed at some point in the position/momentum phase space and we can now construct the trajectories using the aforementioned rules.
As it should be clear from the explicit example of Fig. \ref{figSM1}, no bound states are allowed for a repulsive bell-shaped potential for $v\ne 0$ (differently from the case at $v=0$ [\ocite{wigners5}]).
The same method can be used to study the classical scattering processes, that will lead to determining the LQSS in the semiclassical approximation: for a detailed discussion we refer directly to Fig. \ref{figSM2} and the relative caption.
In general, whether a particle is reflected or transmitted it depends on the height of the potential barrier, but there exists a window of momenta 
such that a left mover (we assume the defect velocity to be positive $v>0$) is unavoidably transmitted. This classically purely transmissive window of momenta encloses 
the corresponding quantum counterpart, but actually is much larger than the latter: this because several channels, while energetically allowed, are classically forbidden because of the absence of tunneling.
Consistently with the quantum case, when $v$ is increased more and more particles are surely transmitted until the sound velocity is reached: beyond that point, all the particles are unavoidably transmitted.

\subsection{The semiclassical LQSS}

Here we derive the semiclassical expression for the LQSS, it is convenient to briefly recap the different contributions
\begin{enumerate}
\item The particles initially sat on the defect are ignored, since contribute as a transient vanishing in time.
\item Left movers that are initially placed on the left of the defect of course never meet the latter and thus do not scatter. Symmetrically, right movers that are initially on the right of the defect do not undergo a scattering process.
\item Left movers initially placed on the right of the defect, as well as right movers coming from the left of the defect, undergo a scattering event. Let $k$ be the momentum of the incoming particle, we denote as $k^s$ the momentum of the scattered particle, the latter being determined through the rules previously described.
\item Along the time evolution, the phase space measure is conserved (Liouville Theorem) thus $\rho_t(x(t),k(t))dx(t)dk(t)=\rho_{t=0}(x,k)dxdk$ where $(x(t),k(t))$ are the time-evolved coordinates with initial conditions $(x,k)$.
After the scattering event has occurred, the time evolution is simply 
\be
x(t)=v(k^s)\left(t-t_{\delta}(k)-\left|\frac{x}{v(k)}\right|\right)\, , \hspace{2pc}k(t)=k^{s}\, .
\ee
were $t_\delta(k)$ is the \emph{delay time} experienced by the particle undergoing the scattering event. As it is clear, the Jacobian of the change of coordinate $(x(t),k(t))\to (x,k)$ simply amounts to compute $d k^s/dk$ . Using the conservation of energy similarly to Eq. (\ref{jacob}) and the Liouville Theorem
\be
\rho_t(x(t),k(t))=\rho_{t=0}(x,k)\left|\frac{v(k)}{v(k^s)}\right| \label{liouville}
\ee
\end{enumerate}

The corrections to the initial ensemble given by the emergent LQSS are most easily described in terms of \emph{holes} and \emph{occupancies}. The holes describe the removal of particles from 
their unperturbed trajectories due to the scattering process on the defect: the front of propagating holes at a given momentum $k$ can reach at most a position $x=v(k) t$. On the other hand, the occupancies describe the phase space populated after the scattering event: the front of particles that undergoes the scattering $k\to k^s$ can travel at most up to $x=v(k^s) t$ (here we neglect the time delay, since it is irrelevant in the large time limit and thus for the determination of the LQSS).
These considerations, together with the fact that the two point correlator $\mathcal{C}_t(x,s)=\langle \tilde{c}^\dagger_{x+s/2}\tilde{c}_{x-s/2}\rangle_t$ is the Fourier transform of the Wigner function (Eq. (10) of the main text), provide us the two point correlator in the large time limit and outside of the defect support
\be
\mathcal{C}_t(x,s)\simeq\int_{-k_f}^{k_f} \frac{dk}{2\pi}\left(1-\theta(v(k))\theta(|v(k)|-\zeta)\right)e^{-ik s}+\theta(v(k^s))\theta(|v(k^s)|-\zeta)\left|\frac{v(k)}{v(k^s)}\right|e^{-ik^{s} s}\, ,
\ee
where, as usual, $\zeta=x/t$. As expected, the semiclassical LQSS coincides with the quantum LQSS in the case where only a scattering channel is possible, i.e. when we completely neglect the tunneling through non-classical trajectories. 

\subsection{The semiclassic correlator on a supersonic defect}

In the semiclassical regime, as well as in the true quantum case, a supersonic defect cannot produce a LQSS: in this case the non-trivial two point correlator that persists at late times must be localized on the defect. 
Deriving such a correlator in the purely quantum case would require knowing the eigenfunctions $\psi_k(x)$ \emph{on the defect}: while in principle possible for the exactly solvable case of the $\delta-$defect, it still involves some rather tedious calculations. Instead, 
a simple and intuitive derivation can be outlined within the semiclassical approximation. 
In fact, the defect sees a flux of incoming particles (from the far right, assuming a positive velocity $v$): at infinity, the energy of the particles is simply equal to their kinetic energy.
While a particle travels across the defect, its momentum is unambiguously determined by energy conservation (in the supersonic case, turning points are absent). Let $q_{k,x}$ be the momentum that the particle has at point $x$ if it had momentum $k$ far away from the defect, thus $q_{k,x}$ is determined by energy conservation
\be
-\cos(q_{k,x})-v q_{k,x}+V(x)=-\cos(k)-vk\, .
\ee
Along the motion the phase space changes as in (\ref{liouville}) with the only difference of using $q_{k,x}$ rather than $k^s$. 
Thus, we finally obtain our two point correlator for a supersonic defect in the infinite time limit
\be
\lim_{t\to\infty}\mathcal{C}_t(x,s)=\int_{-k_f}^{k_f} \frac{dk}{2\pi}\left|\frac{v(k)}{v(q_{k,x})}\right|e^{-iq_{k,x} s}\, .
\ee
which is the result presented in the main text. 
In principle, a similar semiclassical analysis could be pursued even for subsonic defects, however extra complications arise due the presence of turning points and thus multiple solutions to the energy conservation must be considered.


\begin{thebibliography}{99}


\bibitem{exp1}M. Greiner \emph{et al}, 
Nature \href{\doi10.1038/nature00968}{\bf 419}, 51-54 (2002).
\bibitem{exp2}T. Kinoshita, T. Wenger,  and D. S. Weiss, 
 Nature \href{\doi10.1038/nature04693}{\bf 440}, 900 (2006).
 \bibitem{exp3}S. Hofferberth, I. Lesanovsky \emph{et al}, 
Nature \href{\doi10.1038/nature06149}{\bf 449}, 324-327 (2007). 
\bibitem{exp4}L. Hackermuller, U. Schneider \emph{et al}, 
Science \href{\doi10.1126/science.1184565}{\bf 327}, 1621 (2010).
\bibitem{exp5}S. Trotzky, Y.-A. Chen \emph{et al}, 
Nature Phys. \href{\doi10.1038/nphys2232}{\bf 8}, 325 (2012).
\bibitem{exp6}M. Gring, M. Kuhnert \emph{et al}, 
Science \href{\doi10.1126/science.1224953}{\bf 337}, 1318 (2012).
\bibitem{exp7}M. Cheneau, P. Barmettler \emph{et al}, 
Nature \href{http://dx.doi.org/10.1038/nature10748}{\bf 481}, 484 (2012).
\bibitem{exp8}T. Langen, R. Geiger \emph{et al}, 
Nature Physics \href{http://dx.doi.org/10.1038/nphys2739}{\bf 9}, 640 (2013).
\bibitem{exp9}F. Meinert, M.J. Mark \emph{et al}, 
Phys. Rev. Lett. \href{http://dx.doi.org/10.1103/PhysRevLett.111.053003}{\bf 111}, 053003 (2013).
\bibitem{exp10}J.P. Ronzheimer, M. Schreiber \emph{et al}, 
Phys. Rev. Lett. \href{\doi10.1103/PhysRevLett.110.205301}{\bf 110}, 205301 (2013).
\bibitem{exp11} L. Vidmar, J. P. Ronzheimer, M. Schreiber, S. Braun, S. S. Hodgman, S. Langer, F. Heidrich-Meisner, I. Bloch, and U. Schneider
Phys. Rev. Lett. \href{\doi10.1103/PhysRevLett.115.175301}{\bf 115}, 175301 (2015).

\bibitem{newExp1}I. Bloch, Nature Physics \href{\doi 10.1038/nphys138}{\bf 1},23-30 (2005).

\bibitem{newExp2} B. Paredes, Q. Widera, V. Murg, O. Mandel, S. F\"olling, I. Cirac, G. V. Shlyapnikov, T. W. H\"ansch, I. Bloch, Nature \href{\doi 10.1038/nature02530}{\bf 429}, 277-281 (2004)

\bibitem{newExp3} R. J\"ordens, N. Strohmaier, K. G\"unter, H. Moritz, T. Esslinger, Nature \href{\doi 10.1038/nature07244}{\bf 455}, 204-207 (2008).


\bibitem{newExp4}S. Palzer, C. Zipkes, C. Sias, M. K\"ohl, Phys. Rev. Lett. \href{https://doi.org/10.1103/PhysRevLett.103.150601}{\bf 103}, 150601 (2009).

\bibitem{newExp5} T. Fukuhare \emph{et al.}, Nature Physics \href{\doi10.1038/nphys2561}{\bf 9}, 235–241 (2013).


\bibitem{calabrese-cardy}
P. Calabrese and  J. Cardy,
Phys. Rev. Lett. \href{\doi10.1103/PhysRevLett.96.136801}{\bf 96}, 136801 (2006);  
J. Stat. Mech. (2007) \href{\doi10.1088/1742-5468/2007/06/P06008}{P06008}.

\bibitem{pssv}
A. Polkovnikov, K. Sengupta, A. Silva, and M. Vengalattore, 
Rev. Mod. Phys. \href{\doi10.1103/RevModPhys.83.863}{\bf 83}, 863 (2011).

\bibitem{difrancesco} P. Di Francesco, P.  Mathieu, D. S\'en\'echal \emph{Conformal field theory} (Springer Science \& Business Media, 2012).

\bibitem{Korepin} V.E. Korepin, N.M. Bogoliubov, A.G. Izergin, \emph{Quantum Inverse Scattering Method and Correlation Functions} (University Press, Cambridge, 1993).

\bibitem{smirnov}  F. A. Smirnov, \emph{Form factors in completely integrable models of quantum field theory }(World Scientific, 1992).







\bibitem{ggew1}F.H.L. Essler and M. Fagotti, J. Stat. Mech. (2016) \href{\doi10.1088/1742-5468/2016/06/064002}{064002}  .
\bibitem{ggew2}J.-S. Caux and F.H.L. Essler, Phys. Rev. Lett. \href{http://dx.doi.org/10.1103/PhysRevLett.110.257203}{\bf 110}, 257203 (2013).
\bibitem{ggew3}B. Pozsgay, J. Stat. Mech. (2014) \href{\doi10.1088/1742-5468/2014/10/P10045}{P10045} .
\bibitem{ggew4}P. Calabrese, F. H. L. Essler and M. Fagotti,  Phys. Rev. Lett. \href{http://dx.doi.org/10.1103/PhysRevLett.106.227203}{\bf 106}, 227203 (2011).
\bibitem{ggew6}F. H. L. Essler, S. Evangelisti, and M. Fagotti, Phys. Rev. Lett.  \href{http://dx.doi.org/10.1103/PhysRevLett.109.247206}{\bf 109}, 247206 (2012).
\bibitem{ggew9} J.-S. Caux and R. M. Konik, Phys. Rev. Lett. \href{http://dx.doi.org/10.1103/PhysRevLett.109.175301}{\bf 109}, 175301 (2012).
\bibitem{ggew5}M. Fagotti and F.H.L. Essler, Phys. Rev. B \href{http://dx.doi.org/10.1103/PhysRevB.87.245107}{\bf 87}, 245107 (2013).
\bibitem{ggew7} B. Pozsgay, J. Stat. Mech. (2013) \href{http://dx.doi.org/10.1088/1742-5468/2013/07/P07003}{P07003}.
\bibitem{ggew8}M. Fagotti and F. H. L. Essler, J. Stat. Mech. (2013) \href{http://dx.doi.org/10.1088/1742-5468/2013/07/P07012}{P07012}.
\bibitem{ggew10} G. Mussardo, Phys. Rev. Lett. \href{http://dx.doi.org/10.1103/PhysRevLett.111.100401} {\bf 111}, 100401 (2013).

\bibitem{specialissue} Special issue on {\em Quantum Integrability in Out of Equilibrium Systems}, editors P. Calabrese, F.H.L. Essler and G. Mussardo, 
J. Stat. Mech. (2016) 064001.




\bibitem{gge1} M. Rigol, V. Dunjko, V. Yurovsky, and M. Olshanii,
Phys. Rev. Lett. \href{http://dx.doi.org/10.1103/PhysRevLett.98.050405}{\bf 98}, 050405  (2007).


\bibitem{gge3}M. Rigol, Phys. Rev. Lett. \href{\doi10.1103/PhysRevLett.103.100403}{\bf 103}, 100403 (2009).

\bibitem{ggef1}B. Pozsgay, M. Mesty\'{a}n , M. A. Werner, M. Kormos, G. Zar\'{a}nd, and G. Tak\'{a}cs, Phys. Rev. Lett. \href{http://dx.doi.org/10.1103/PhysRevLett.113.117203}{{\bf 113}}, 117203 (2014).
\bibitem{ggef2}M. Mestyan, B. Pozsgay, G. T\'akacs, M.A. Werner, J. Stat. Mech. {2015} \href{\doi10.1088/1742-5468/2015/04/P04001}{P04001}.
\bibitem{ggef3}J. De Nardis, B. Wouters, M. Brockmann, and J.-S. Caux, Phys. Rev. A \href{http://dx.doi.org/10.1103/PhysRevA.89.033601}{\bf 89}, 033601 (2014).
\bibitem{ggef4}B. Wouters, J. De Nardis, M. Brockmann, D. Fioretto, M. Rigol, and J.-S. Caux, Phys. Rev. Lett.  \href{http://dx.doi.org/10.1103/PhysRevLett.113.117202}{\bf 113}, 117202 (2014).




\bibitem{lch1} E. Ilievski, J. De Nardis, B. Wouters, J-S Caux, F. H. L. Essler, T. Prosen, Phys. Rev. Lett. \href{\doi10.1103/PhysRevLett.115.157201}{\bf 115}, 157201 (2015). 
\bibitem{lch2}E. Ilievski, M. Medenjak, and T. Prosen, Phys. Rev. Lett. \href{http://dx.doi.org/10.1103/PhysRevLett.115.120601}{\bf 115}, 120601 (2015).
\bibitem{lch3}E. Ilievski, M. Medenjak \emph{et al}, J. Stat. Mech. (2016) \href{\doi10.1088/1742-5468/2016/06/064008}{064008}.
\bibitem{lch4}E. Ilievski, E. Quinn, J. De Nardis, M. Brockmann J. Stat. Mech. (2016) \href{\doi10.1088/1742-5468/2016/06/063101}{P063101}.
\bibitem{lch5}L. Piroli, E. Vernier, J. Stat. Mech. (2016) \href{\doi10.1088/1742-5468/2016/05/053106}{P053106}.
\bibitem{lch6} L. Piroli, E. Vernier, P. Calabrese, Phys. Rev. B \href{\doi10.1103/PhysRevB.94.054313}{\bf 94}, 054313 (2016).

\bibitem{lchf1}S. Sotiriadis, Phys. Rev. A \href{https://doi.org/10.1103/PhysRevA.94.031605}{\bf 94}, 031605 (2016).
\bibitem{lchf2}F. H. L. Essler, G. Mussardo, and M. Panfil,  Phys. Rev. A \href{http://dx.doi.org/10.1103/PhysRevA.91.051602}{\bf 91}, 051602(R) (2015).
\bibitem{lchf3}A. Bastianello, S. Sotiriadis, J. Stat. Mech \href{\doi10.1088/1742-5468/aa5738 }{023105} (2017).
\bibitem{lchf4} F. H. L. Essler, G. Mussardo, M. Panfil, J. Stat. Mech. (2017) \href{http://stacks.iop.org/1742-5468/2017/i=1/a=013103}{013103}.
\bibitem{lchf5} E. Vernier and A. C. Cubero, J. Stat. Mech. (2017) \href{http://stacks.iop.org/1742-5468/2017/i=2/a=023101}{023101}.


\bibitem{liebrob} E.H. Lieb, D.W. Robinson, Commun. Math. Phys.\href{http://dx.doi.org/10.1007/BF01645779}{\bf 28}, 251 (1972).
See also 
B. Bertini, Phys. Rev. B \href{https://link.aps.org/doi/10.1103/PhysRevB.95.075153}{95, 075153} (2017), for an analysis of 
Galilean systems.

\bibitem{localquenches}
A. De Luca, Phys. Rev. B \href{\doi10.1103/PhysRevB.90.081403}{\bf 90} 081403 (2014);
B. Bertini and M. Fagotti, Phys. Rev. Lett. \href{\doi10.1103/PhysRevLett.117.130402}{\bf 117} 130402 (2016),
M. Fagotti, \href{https://arxiv.org/abs/1508.04401}{arXiv:1508.04401} (2015).

\bibitem{transportbertini}B. Bertini, M. Collura, J. De Nardis, M. Fagotti, Phys. Rev. Lett. \href{\doi10.1103/PhysRevLett.117.207201}{\bf 117}, 207201 (2016).

\bibitem{NESSnum1} T. Sabetta, G. Misguich, Phys. Rev. B \href{https://doi.org/10.1103/PhysRevB.88.245114}{\bf 88}, 245114  8 (2013).
\bibitem{NESSnum2} C. Karrasch, R. Ilan, and J. E. Moore, Phys. Rev. B\href{https://doi.org/10.1103/PhysRevB.88.195129}{\bf  88}, 195129 (2013).
\bibitem{NESSnum3} A. Biella, A. De Luca, J. Viti, D. Rossini, L. Mazza,  R. Fazio
Phys. Rev. B \href{https://doi.org/10.1103/PhysRevB.93.205121}{\bf 93}, 205121 (2016).



\bibitem{NESSf0} V. Eisler, Z. R\'acz, Phys. Rev. Lett. \href{https://doi.org/10.1103/PhysRevLett.110.060602}{\bf 110}, 060602  (2013)
\bibitem{NESSf1}Y. Ogata, Phys. Rev. E \href{\doi10.1103/PhysRevE.66.066123}{\bf66} (6), 066123 (2002).
\bibitem{NESSf2}Y. Ogata, Phys. Rev. E \href{\doi10.1103/PhysRevE.66.016135}{\bf 66}(1), 016135 (2002).
\bibitem{NESSf3}D. Karewski, European Phys. J. B-Cond. Mat. \href{\doi10.1140/epjb/e20020139}{\bf 27}(1), 147 (2002).
\bibitem{NESSf4}T. Platini and D. Karevski, European Phys. J. B \href{\doi10.1140/epjb/e2005-00402-2}{\bf48}(2),225 (2005).
\bibitem{NESSf5}A. De Luca, J. Viti, D. Bernard, B. Doyon, Phys. Rev. B \href{\doi10.1103/PhysRevB.88.134301}{\bf88}(13), 1342301 (2013).
\bibitem{NESSf6}A. De Luca, G. Martelloni, J. Viti, Phys. Rev. A \href{\doi10.1103/PhysRevA.91.021603}{\bf91}(2), 021603 (2014). 
\bibitem{NESSf7} B. Doyon, A. Lucas, K. Schalm, M. J. Bhaseen, J. Phys. A: Math. Theor, \href{https://doi.org/10.1088/1751-8113/48/9/095002}{\bf 48} 095002 (2015).
\bibitem{NESSf8}J. Viti, J.-M. St\'ephan, J. Dubail, M. Haque, EPL (Europhysics Letters) \href{\doi10.1209/0295-5075/115/40011}{\bf 115}(4), 40011 (2016).
\bibitem{NESSf11} L. Vidmar, D. Iyer, M. Rigol Phys. Rev. X 7,\href{\doi10.1103/PhysRevX.7.021012}{021012} (2017).
\bibitem{NESSf12} L. Vidmar, W. Xu, M. Rigol, \href{https://arxiv.org/abs/1704.01125}{arXiv:1704.01125} (2017).
\bibitem{NESSf9}M. Kormos, \href{https://arxiv.org/abs/1704.03744}{arXiv:1704.03744} (2017).
\bibitem{NESSf10} G.Perfetto, A. Gambassi \href{https://arxiv.org/abs/1704.03437}{arXiv:1704.03437}(2017).



\bibitem{NESSc1}S. Sotiriadis, J. Cardy, J. Stat. Mech. \href{\doi10.1088/1742-5468/2008/11/P11003}{P11003} (2008).
\bibitem{NESSc2} D. Bernard, B. Doyon, J. Phys. A: Math. Theor. \href{https://doi.org/10.1088/1751-8113/45/36/362001}{\bf45} 362001 (2012).
\bibitem{NESSc3} M. J. Bhaseen, B. Doyon, Nature Physics \href{\doi10.1038/nphys3320}{\bf 11} 509-514 (2015).
\bibitem{NESSc4} D. Bernard, B. Doyon, Ann. Henri Poincar\'e (2015) \href{\doi10.1007/s00023-014-0314-8}{16:113}.
\bibitem{NESSc5}D. Bernard, B. Doyon, J. Stat. Mech. (2016) \href{https://doi.org/10.1088/1742-5468/2016/03/033104}{033104}.
\bibitem{PB} B. Bertini, L. Piroli, P. Calabrese, arxiv: \href{https://arxiv.org/abs/1709.10096}{1709.10096}.





\bibitem{NESSI1}C. Karrasch, R. Ilan, J. E. Moore, Phys. Rev. B \href{\doi10.1103/PhysRevB.88.195129}{\bf 88}(19), 195129 (2013)
\bibitem{NESSI2} A. De Luca, J. Viti, L. Mazza, D. Rossini, Phys. Rev. B \href{https://journals.aps.org/prb/abstract/10.1103/PhysRevB.90.161101}{\bf 88}(19), 195129 (2013).
\bibitem{NESSI3}A. De Luca, M. Collura, J. De Nardis, \href{https://arxiv.org/abs/1612.07265v2}{arXiv:1612.07265v2} (2016).

\bibitem{hydrodoyon1} O. A. Castro-Alvaredo, B. Doyon, T. Yoshimura, Phys. Rev. X \href{\doi10.1103/PhysRevX.6.041065}{\bf 6}, 041065 (2016).
\bibitem{PBCF}L. Piroli, J. De Nardis, M. Collura, B. Bertini, M. Fagotti, Phys. Rev. B\href{\doi 10.1103/PhysRevB.96.115124}{\bf 96}, 115124 (2017).

\bibitem{hydrodoyon2} B. Doyon, T. Yoshimura \href{\doi: 10.21468/SciPostPhys.2.2.014}{SciPost Phys. 2, 014} (2017).
 
\bibitem{IlNard} E. Ilievski, J. De Nardis, \href{https://arxiv.org/abs/1702.02930}{arXiv:1702.02930}.
\bibitem{new_doyon1} B. Doyon, J. Dubail, R. Konik, T. Yoshimura, \href{https://arxiv.org/abs/1704.04151}{arXiv:1704.04151}.
\bibitem{new_doyon2} B. Doyon, T. Yoshimura, J.-S. Caux \href{https://arxiv.org/abs/1704.05482}{arXiv:1704.05482}.
\bibitem{xxzhydro} V. B. Bulchandani, R. Vasseur, C. Karrasch, J. E. Moore, \href{https://arxiv.org/abs/1704.03466}{ arXiv:1704.03466 }(2017).
\bibitem{karrasch} Vir B. Bulchandani, R. Vasseur, C. Karrasch, J. E. Moore 	\href{https://arxiv.org/abs/1702.06146}{arXiv:1702.06146} (2017).
\bibitem{doyonspohn} B. Doyon, H. Spohn, 	\href{https://arxiv.org/abs/1705.08141}{arXiv:1705.08141} (2017).


\bibitem{hardcore} L. Tonks, Phys. Rev. \href{http://dx.doi.org/10.1063/1.1703687}{\bf 50}, 955 (1936), M. Girardeau, J. Math. Phys. \href{http://dx.doi.org/10.1063/1.1703687}{\bf 1}, 516 (1960).


\bibitem{refB1} I. Hans, J. Stockhofe, P. Schmelcher, Phys, Rev. A. \href{\doi 10.1103/PhysRevA.92.013627}{\bf 92}, 013627 (2015).
\bibitem{refB2} M. Cheng, V. Galitski, A. Das Sarma, Phys. Rev. B \href{\doi 10.1103/PhysRevB.84.104529}{\bf 84}, 104529 (2011).
\bibitem{refB3} T. Karzig, G. Refael, F. von Oppen, Phys. Rev. X \href{\doi 10.1103/PhysRevX.3.041017}{\bf 3}, 041017 (2013).
\bibitem{refB4} T. Karzig, A. Rahmani, F. von Oppen, G. Refael, Phys. Rev, B \href{\doi 10.1103/PhysRevB.91.201404}{\bf 91}, 201404(R) (2015).

\bibitem{imp1} A. H. Castro Neto, M. P. A. Fisher, Phys. Rev. B \href{https://doi.org/10.1103/PhysRevB.53.9713}{\bf 53 }, 9713 (1996).

\bibitem{imp2} M. Schecter, A. Kamenev, D. M. Gangardt, A. Lamacraft, Phys. Rev. Lett. \href{https://doi.org/10.1103/PhysRevLett.108.207001}{\bf 108}, 207001 (2012).

\bibitem{imp3} M. B. Zvonarev, V. V. Cheianov, T, Giamarchi, Phys. Rev. Lett. \href{\doi10.1103/PhysRevLett.99.240404}{\bf 99}, 240404 (2007).

\bibitem{imp4} A. Kamenev, L. I. Glazman, Phys. Rev. A \href{\doi10.1103/PhysRevA.80.011603}{\bf 80}, 011603(R) (2009).

\bibitem{imp5} S. Peotta, D. Rossini, M. Polini, F. Minardi, R. Fazio, Phys. Rev. Lett. \href{\doi10.1103/PhysRevLett.110.015302}{\bf110},015302 (2013).

\bibitem{imp6} M. Schecter, D. M. Gangardt, A.Kamenev, Annals of Physics \href{https://doi.org/10.1016/j.aop.2011.10.001
}{\bf 327} (2012) 639-670.

\bibitem{imp7} A. S. Campbell, D. M. Gangardt, SciPost Phys. \href{https://scipost.org/SciPostPhys.3.2.015}{\bf 3}, 015 (2017).

\bibitem{solutionXY1} E. Lieb, T. Schultz, D. Mattis, Ann. of Phys. \href{\doi10.1016/0003-4916(61)90115-4}{16}, 407-466 (1961).
\bibitem{solutionXY2} Th. Niemeijer, Physica \href{\doi10.1016/0031-8914(67)90235-2}{36}, 377-419 (1967).
\bibitem{solutionXY3}Niemeijer, Physica \href{\doi10.1016/0031-8914(68)90085-2}{39}, 313-326 (1968).






\bibitem{suppl} Supplementary material at [url] with ...

\bibitem{LL1} M. A. Cazalilla, J. Phys. B. \href{http://stacks.iop.org/0953-4075/37/i=7/a=051}{\bf 37} S1 (2004).

\bibitem{LL2} S. Lukyanov, Nucl. Phys. B \href{https://doi.org/10.1016/S0550-3213(98)00249-1}{\bf 522}, 533-549 (1998).

\bibitem{sakurai} J. J. Sakurai, E. D.  Commins, \emph{ Modern quantum mechanics, revised edition} (1995).



\bibitem{gaussification1}S. Sotiriadis, P. Calabrese, J. Stat. Mech. (2014) \href{http://dx.doi.org/10.1088/1742-5468/2014/07/P07024}{P07024}.
\bibitem{gaussification2}S. Sotiriadis, G. Martelloni, J. Phys. A \href{\doi10.1088/1751-8113/49/9/095002}{\bf4 9} 095002 (2016).
\bibitem{gaussification3}M. Gluza, C. Krumnow, M. Friesdorf, C. Gogolin, J. Eisert, Phys. Rev. Lett. \href{\doi10.1103/PhysRevLett.117.190602}{117, 190602} (2016).
\bibitem{gaussification4}S. Sotiriadis, arXiv:\href{https://arxiv.org/abs/1612.00373}{1612.00373}.


\bibitem{wigner1} E. Wigner, Phys. Rev. \href{\doi.org/10.1103/PhysRev.40.749}{\bf 40}, 749 (1932).
\bibitem{wigner2}M. Hillery, R.F. O’Connell, M.O. Scully and E.P. Wigner, Phys. Rep. \href{http://dx.doi.org/10.1016/0370-1573(84)90160-1}{\bf 106} (1984) 121.
\bibitem{wigner3}F.J. Narcowich Physica 134 \href{http://dx.doi.org/10.1016/0378-4371(85)90161-X}{\bf A} (1985) 193-208. 


\bibitem{wigners1} E. Bettelheim, P. Wiegmann, Phys. Rev. B \href{\doi10.1103/PhysRevB.84.085102}{\bf 84}, 085102 (2011).
\bibitem{wigners2}E. Bettelheim, L. Glazman, Phys. Rev. Lett. \href{\doi10.1103/PhysRevLett.109.260602}{\bf 109}, 260602 (2012).
\bibitem{wigners3}I. V. Protopopov, D. B. Gutman, P. Schmitteckert, A. D. Mirlin, Phys, Rev. B \href{\doi10.1103/PhysRevB.87.045112}{\bf87}, 045112(2013).
\bibitem{wigners4} A. Bastianello, M. Collura, S. Sotiriadis, Phys. Rev. \href{https://doi.org/10.1103/PhysRevB.95.174303}{\bf  B} 95, 174303 (2017).
\bibitem{wigners5} P. Wendenbaum, M. Collura, D. Karevski Phys. Rev. A \href{https://doi.org/10.1103/PhysRevA.87.023624}{\bf 87}, 023624 (2013).

\bibitem{refA1}E. Bettelheim, L. Glazman, Phys. Rev. Lett. \href{\doi 10.1103/PhysRevLett.109.260602}{\bf 109}, 260602 (2012).
\bibitem{refA2} I. V. Propotov, D B. Gutman, P. Schmitteckert, A. D. Mirlin, Phys. Rev. B \href{\doi 10.1103/PhysRevB.87.045112}{\bf 87},045112 (2013).
\bibitem{refA3}I. V. Propotov, D B. Gutman, M. Oldenburg, A. D. Mirlin, Phys. Rev. B \href{\doi 10.1103/PhysRevB.89.161104}{\bf 89},161104(R) (2014).
\bibitem{F17} M. Fagotti, 	arXiv:\href{https://arxiv.org/abs/1708.05383}{1708.05383}.


\bibitem{curvecft} J. Dubail, J.-M. St\'ephan, J. Viti, P. Calabrese  \href{https://doi.org/10.21468/SciPostPhys.2.1.002}{SciPost Phys. 2, 002 (2017)} .






\end{thebibliography}
\end{document}